\documentclass[acmsmall,nonacm]{acmart}
\renewcommand\footnotetextcopyrightpermission[1]{}
\AtBeginDocument{%
  }

\setcopyright{none}

\usepackage{xspace}     
\usepackage{xcolor}     
\usepackage{nicefrac}   
\usepackage{amsmath}    
\usepackage{letltxmacro} 
\usepackage{colortbl}
\usepackage{subcaption}
\usepackage{listings}
\newcommand{\figref}[1]{Figure~\ref{fig:#1}}
\newcommand{\secref}[1]{Section~\ref{sec:#1}}

\newcommand{\appref}[1]{Appendix~\ref{app:#1}}
\newcommand{\tableref}[1]{Table~\ref{tab:#1}}

\newcommand{\etal}{et al.\xspace}

\newcommand{\cdms}{\,cd/m$^2$\xspace}
\newcommand{\ppd}{\,ppd\xspace}

\newcommand{\jod}{\,JOD\xspace}
\newcommand{\mbps}{\,Mbps\xspace}

\LetLtxMacro{\originaleqref}{\eqref}
\renewcommand{\eqref}[1]{Eq.~\originaleqref{eq:#1}}

\newcommand{\revision}[1]{{#1}}
\newcommand{\edit}[1]{{#1}}

\begin{document}

\title{Streaming across viewing conditions --- The effect of viewing distance and display luminance on video quality}

\author{Dounia Hammou}
\email{dh706@cl.cam.ac.uk}
\orcid{0000-0001-7475-6722}
\affiliation{%
  \institution{University of Cambridge}
  \city{Cambridge}
  \country{UK}
}
\author{Luk\'{a}\v{s} Krasula}
\email{l.krasula@gmail.com}
\affiliation{%
  \institution{Netflix, Inc.}
  \city{Los Gatos}
  \state{CA}
  \country{USA}
}
\author{Christos G. Bampis}
\email{christosb@netflix.com}
\affiliation{%
  \institution{Netflix, Inc.}
  \city{Los Gatos}
  \state{CA}
  \country{USA}
}
\author{Zhi Li}
\email{zli@netflix.com}
\affiliation{%
  \institution{Netflix, Inc.}
  \city{Los Gatos}
  \state{CA}
  \country{USA}
}
\author{Rafa{\l} K. Mantiuk}
\email{rafal.mantiuk@cl.cam.ac.uk}
\orcid{0000-0003-2353-0349}
\affiliation{%
  \institution{University of Cambridge}
  \city{Cambridge}
  \country{UK}
}

\renewcommand{\shortauthors}{Hammou et al.}

\begin{abstract}
The visibility of streamed video distortions is affected by the user's display and viewing conditions, such as their display's luminance or the viewing distance. Distortions become less visible at lower luminance levels or farther viewing distances. However, most image and video quality metrics assume fixed optimal display and viewing conditions, making them insensitive to the viewing conditions. In this work, we demonstrate how existing metrics can be adapted to account for both factors and compare such an adaptation with the metrics that natively model viewing conditions. First, we collected a new video quality dataset, HDR-VDC, which contains streaming distortions in 4K HDR videos. 
The dataset consists of 12 4K and 4 FHD HDR source video content, each distorted with AV1 encoding and Lanczos upscaling from a lower resolution. The quality loss due to the distortions was measured at two viewing distances, corresponding to 60 and 120 pixels per visual degree, and two display mean luminance levels, varying by a factor of 8 in terms of luminance. 
Next, we developed a pre-processing framework that lets us adapt many quality metrics so that they can evaluate content across the viewing distances and luminance levels. The framework employs a display model and perceptually uniform encoding, PU21, to account for the effect of display luminance, and perspective projection scaling (PPS) to account for the effect of viewing distance. We used the framework to compare the performance of 35 state-of-the-art metrics. This allowed us to identify a selection of existing quality metrics that can be effectively adapted to account for the viewing conditions. 
The dataset is available at \href{https://doi.org/10.17863/CAM.107964}{\textcolor{blue}{https://doi.org/10.17863/CAM.107964}} and the code at \href{https://github.com/gfxdisp/HDR-VDC}{\textcolor{blue}{https://github.com/gfxdisp/HDR-VDC}}.
\end{abstract}





\begin{teaserfigure}
    \includegraphics[width=\textwidth]{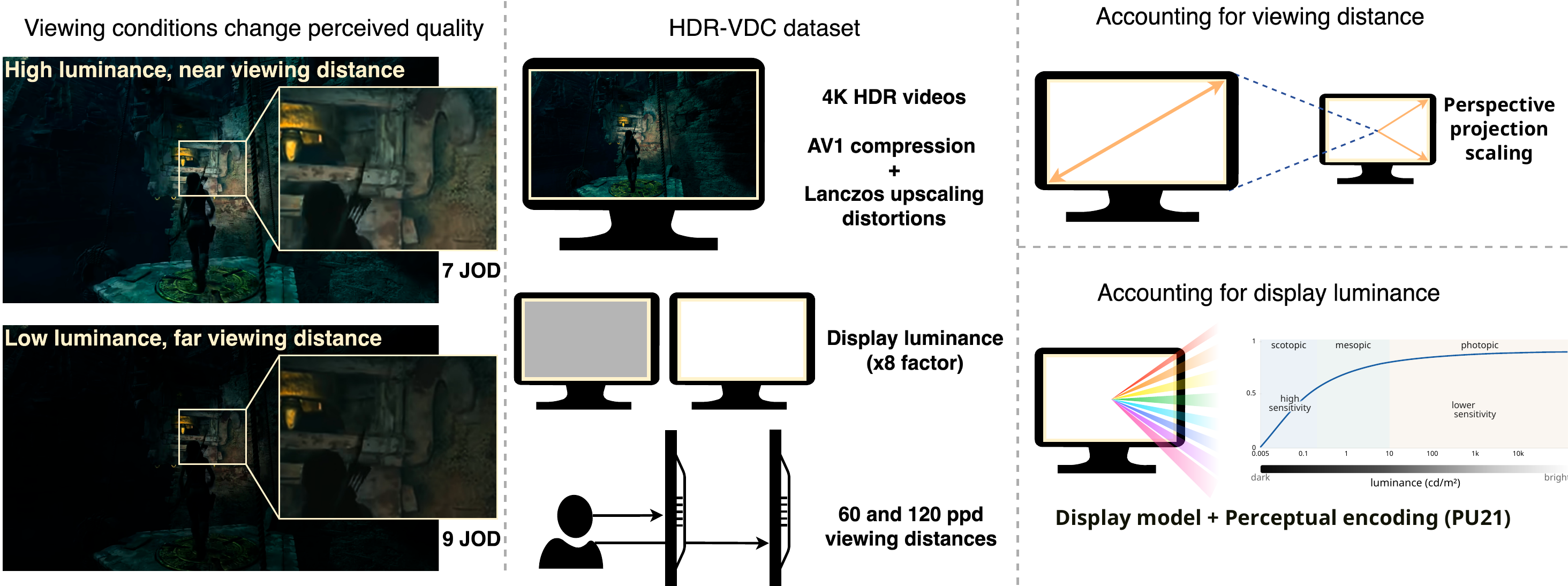}
    \caption{The visibility of streaming distortions depends on viewing conditions. The same compressed 4K HDR content exhibits fewer visible artifacts at lower display luminance or greater viewing distances. We introduce HDR-VDC, a dataset that captures these effects across luminance levels and viewing distances, and evaluate whether existing quality metrics can be adapted using a display model, perceptual encoding, and perspective projection scaling. Our results show that, while some metrics can account for these factors, many remain unreliable, underscoring the need for quality metrics that natively model real-world viewing conditions.}
\end{teaserfigure}

\maketitle

\section{Introduction}

Video quality metrics play a key role in optimizing the video streaming pipelines. They can, for example, guide the selection of the optimal encoding parameters, ensuring the highest possible quality within the client’s network bandwidth constraints \cite{katsavounidis2018dynamic}. However, while such optimization accounts for the network bandwidth, it overlooks the display and viewing conditions of the client, such as the display size, the display luminance, and the distance from which the video is viewed. Such factors substantially affect the perceptual video quality \cite{amirpour2021impact, sugito2022modeling, barman2023subjective, mikhailiuk2021effect}.
Distortions become less noticeable at greater viewing distances and lower luminance levels --- both effects explained by the contrast sensitivity function (CSF) \cite{Barten1999, hammou2024hvei, ashraf2024castlecsf}.

The effect of viewing distance is primarily modeled in quality metrics using contrast sensitivity functions (CSFs)  \cite{andersson2020flip, zhang1996spatial} or visual difference predictors \cite{mantiuk2023hdr, mantiuk2021fovvideovdp, mantiuk2024colorvideovdp}. 
The CSF models the smallest contrast detectable by an average observer on a uniform background. It is a function of spatial frequency, temporal frequency, luminance, color, and other factors \cite{Barten1999,ashraf2024castlecsf}. Alternative methods have been developed to adapt the existing quality metrics to the varying luminance levels \cite{Mantiuk_Azimi_2021,aydin2008extending} and viewing distances \cite{gu2013self, hammou2024hvei, ye2019predicting}. However, due to the limited availability of datasets measuring these effects, comparing metrics or training them for the task of predicting the quality across viewing distance and display's luminance levels have been proven to be challenging \cite{hammou2023comparison}.

Hence, in this paper, we investigate whether existing quality metrics can be adapted to account for both effects using the combination of previously proposed techniques: PU21 encoding \cite{Mantiuk_Azimi_2021} and PPS rescaling. Since no existing video quality dataset includes both effects, we collect a new dataset designed to study HDR video streaming quality under varying Display and Viewing Conditions, which will be referred to as HDR-VDC. Table I of the supplementary summarizes the main differences between HDR-VDC and existing datasets.
In this dataset, we focus solely on the visibility of distortions as a measure of quality \cite{Bampis2023measuring}, as our goal is to develop quality metrics to optimize video streaming at each display and viewing condition.
The dataset captures the visibility of AV1 compression and Lanczos3 upscaling artifacts --- two major distortions in video streaming applications --- at two viewing distances: 1.1\,m and 2.2\,m, equivalent to 60\ppd and 120\ppd (pixels per degree), and two display luminance levels that differed by 3 stops (a factor of 8). 
The dataset features modern 4K HDR video content, both camera-captured (natural scenes) and computer-generated (animated movies and video games). In contrast to existing datasets that measured quality across multiple viewing distances using direct rating protocols \cite{amirpour2021impact, sugito2022modeling,Bampis2023measuring}, we employed the pairwise comparison (PC) protocol \cite{perez2017practical} with active sampling (ASAP \cite{mikhailiuk2021active}). This approach allows for more accurate and reliable measurement of the effect of viewing distances and display luminance levels, notably when these effects are subtle. Moreover, in comparison to previous studies employing the PC to study the impact of viewing distance \cite{keller2023influence, kufa2019visual, mikhailiuk2021effect}, we are the first to conduct cross-viewing distance comparisons, comparing test videos at two different viewing distances simultaneously. This method enhances the accuracy of assessing the impact of viewing distance and display's luminance on distortions visibility. 

The HDR-VDC was initially introduced in our QoMEX conference paper \cite{hammou2024hdr_vdc}. In this journal paper, we provide a more detailed explanation of the experimental procedure,  analyze the statistical significance of the viewing and display conditions per content and distortion, and compare our dataset to existing ones. Furthermore, we use the dataset to benchmark 35 state-of-the-art metrics, some natively capable of modeling the effect of viewing distance and luminance, some adapted to the task using the combination of PU21 encoding \cite{Mantiuk_Azimi_2021} and PPS rescaling. We also disentangle the effects of viewing conditions from general metric performance to indicate which metrics can be adapted to account for these effects.

\section{Related works}\label{sec:related_works}

\subsection{The effect of viewing distance and display luminance on the visibility of distortions}\label{sec:related_works_vd_luminance}

\citet{liu2014cid}, \citet{gu2015quality}, and \citet{fang2015evaluation} were the first to study the effect of viewing distance on image distortions, notably noise, blur, JPEG, JPEG2000, and gamut mapping artifacts. Following that, \citet{amirpour2021impact} studied this effect on the visibility of H.265 compression distortions on 2K videos. Similarly, \citet{sugito2022modeling} studied it on the visibility of streaming distortions on 8K videos, primarily H.266 compression and upscaling artifacts. All studies employed an absolute category rating (ACR) protocol, and they reported similar findings, in which the visibility of distortions decreases (the perceptual quality of distorted content increases) with the increase in viewing distance. 
\citet{Bampis2023measuring} presented the first crowdsourcing study of the effect of viewing distance on the visibility of H.264 compression and upscaling distortions. The authors simulated the effect of viewing distance by altering the canvas size of the videos and employed the SAMVIQ protocol to measure the visibility of distortions at different canvas sizes. While the methodology differs from previous studies, the authors reported similar findings. 

Others studied the effect of viewing distance on different but related topics. \citet{kufa2019visual} and \citet{keller2024assessing} investigated the effect of viewing distance on the preference for high-resolution videos over lower-resolution ones. 
Unlike previous studies, these employed a 5-scale pairwise comparison protocol, and both reported similar findings, in which the preference for higher-resolution videos decreases as viewing distance increases, indicating that the visibility of upscaling distortions decreases with the increase in viewing distance. 
More broadly, viewing conditions have been studied in different contexts. \citet{gulliver2006defining} investigated how network-, media-, and content-level parameters affect user quality perception in distributed multimedia, and reported that the display type significantly impacts both user information assimilation and perception of quality. \citet{liu2024perceptual} studied how viewing conditions and behaviors --- specifically viewport and head movement --- in VR environments influence the perceptual quality of omnidirectional images, finding that both significantly shape the immersive quality of experience. These findings further support the importance of considering viewing conditions when evaluating perceptual quality and distortion visibility.


The effect of display luminance has mainly been studied in terms of observers’ preferences for display peak luminance \cite{hanhart2014subjective, bist2017qoe}. Prior studies have leveraged these findings to develop methods for automatically adapting display luminance according to image content and viewing conditions, such as ambient illumination \cite{bist2017qoe, schuchhardt2014caped}, as well as for reducing power consumption while preserving perceived visual quality \cite{liu2016content}. However, not many works have investigated the effect of the display luminance on the visibility of distortions.

\citet{mikhailiuk2021effect} were one of the first to measure the effect of the display luminance and viewing distance on the visibility of distortions. The authors measured the visually lossless threshold for which image compression distortions became imperceptible across two viewing distances and two display luminance levels. 
The authors reported an overall decrease in the visually lossless threshold (decrease in the visibility of compression distortions) with the increase in viewing distance or decrease in the display luminance.

Following that, \citet{ebenezer2024hdr} were the first to study the effect of different displays on the quality of compressed and downscaled SDR and HDR videos. The authors employed 3 devices: an OLED TV with a peak luminance of 1028\cdms{}, a QLED TV with a peak luminance of 1170\cdms{}, and an LCD TV with a peak luminance of 230\cdms{}. The authors found that the users' preference for HDR or SDR strongly depended on the display in which they were shown. However, because of the many factors in which the displays differ, including the luminance level, tone mapping, contrast, color gamut, etc.\ it is hard to disentangle the impact of each factor on the visibility of distortions.

In comparison to these works our study is the first, as far as we know, that measures the effect of both viewing distance and display luminance on the visibility of streaming distortions in 4K HDR video content. Furthermore, we employ a pairwise comparison protocol and perform cross-viewing distance comparisons for more accurate measurements of the effect of the display and viewing conditions.

\subsection{HDR video quality assessment datasets}\label{sec:related_works_hdr_datasets}

\citet{vcadik2011evaluation} were the first to evaluate the visibility of HDR distortions in rendered videos, such as noise, compression, luminance modulation, and tone mapping, making it the first publicly available HDR video quality dataset. 
Following this, \citet{azimi2018evaluating}, \citet{rerabek2015subjective}, and \citet{narwaria2015hdr} studied the visibility of distortions due to noise, blur, compression, and luminance shift in camera-captured HDR videos. 
In contrast to these works, \citet{pan2018hdr} and \citet{athar2019perceptual} were the first to study the quality of HDR videos in standard formats (PQ and HLG). Both studies investigated the effect of compression artifacts on 4K HDR videos. Nonetheless, none of these datasets (excluding \cite{vcadik2011evaluation}) were made publicly available.

\citet{shang2022subjective} introduced the first publicly available quality dataset for 4K HDR10 videos, measuring the visibility of streaming distortions (H.265 compression and upscaling artifacts). Subsequently, \citet{shang2023subjective} studied a further range of live-streaming distortions on sports content. However, this dataset has not been made available. Moreover, \citet{ebenezer2024hdr} and \citet{chen2025hdrsdrvqasubjectivevideoquality} compared the visibility of H.265 compression and upscaling artifacts on 4K SDR and HDR videos across different devices. The datasets were made publicly available
\citet{krasula2023subjective} presented the first 4K Dolby Vision dataset collected via crowdsourcing. The authors employed a home study using pre-calibrated LG OLED TVs to control the subjective experiment. Nonetheless, this dataset has also not been made publicly available.
Concurrently, \citet{AVT_VQDB_UHD_1_HDR} presented a new publicly available 4K HDR10 dataset measuring the visibility of multiple compression distortions (H.265, H.266, and AV1) and upscaling artifacts. 

In contrast to the presented datasets, our dataset is the first dataset that measures the effect of display and viewing conditions on the visibility of HDR streaming distortions, as well as one of the first works that studies the visibility of the AV1 compression artifacts on HDR content. A comparison between our dataset and existing datasets is reported in \tableref{datasets_comparison}.

\section{HDR-VDC dataset}\label{sec:subjective_experiment}



\subsection{Source Sequences}\label{sec:source_sequences}

Selecting a diverse set of HDR source content is essential to reliably measure the effect of viewing distance and display luminance on the perceptual quality of streamed HDR videos. Therefore, we collected 50 HDR10, 4K ($3840 \times 2160$) / 2K ($1920 \times 1080$) video sequences from multiple sources \cite{netflixopencontent, barman2021user, song2016sjtu, FroehlichSPIE2014}. These sequences consist of camera-captured, animated, and computer-generated/gaming scenes. 
While our primary focus was to study solely 4K content, we decided to include several full-HD sequences from the color-graded Stuttgart dataset \cite{FroehlichSPIE2014} due to their significantly higher dynamic range. \revision{Moreover, these contents allow us to understand the impact of streaming full-HD videos to clients with a 4K display.}

All sequences are characterized by an SMPTE ST2084 (PQ) electro-optical transfer function (EOTF), ITU-R BT.2020 color space, 10-bit depth, and a YUV\,4:2:0 chroma subsampling. Additionally, each sequence was manually trimmed into a 5 to 11-second video while avoiding abrupt or awkward end cuts that may disturb the viewing experience. \revision{None of the sequences contained audio, to focus the study on the visual aspect alone.} 
Each video sequence was stored in its raw format when calculating the video descriptors outlined below.

Following the ITU-T P.910 recommendation \cite{itu-t-p910}, we calculated spatial and temporal information for each source sequence. However, these standard descriptors alone do not fully capture the range of HDR characteristics. To address this, we incorporated two additional descriptors: dynamic range and median luminance. Details on how each descriptor was calculated can be found in Section 1.1 of the supplementary. 
\begin{figure*}
\centering
\includegraphics[width=\textwidth]{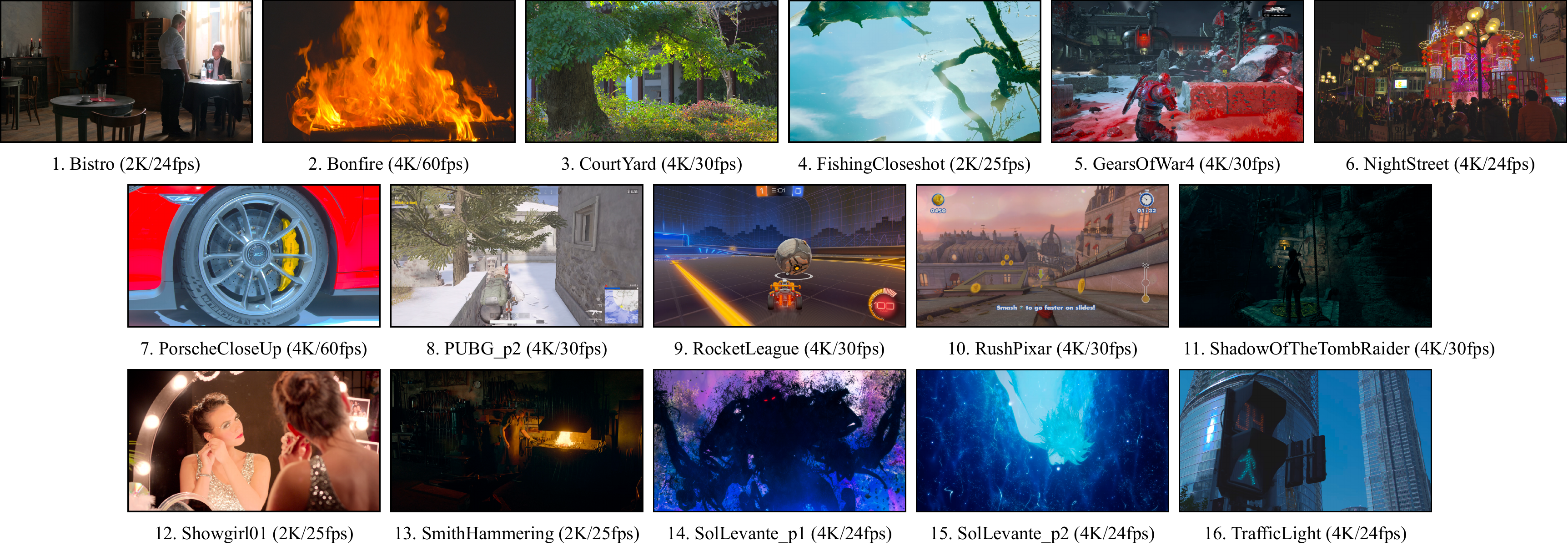}
\caption{Screenshots of the selected reference videos (The sequences are tone-mapped with ffmpeg using the `hable' algorithm \cite{Hable2010FilmicTonemapping} to BT.709 for visualization). For each selected video, we show its resolution and framerate.}
\label{fig:reference_screenshots}
\end{figure*}
The four descriptors were utilized to select 16 reference videos from the 50 collected source sequences. The selection process was carried out using the K-Means algorithm \cite{hartigan1979k}, which clusters the input video descriptors into 16 distinct groups corresponding to the designated number of reference videos.
From each cluster, one video was chosen to ensure a diverse representation of HDR content across spatial information, temporal information, dynamic range, and median luminance. Furthermore, we ensured a balanced number of camera-captured and animated/computer-generated content in the reference video content. 
The thumbnails of the selected 16 reference videos are presented in \figref{reference_screenshots}, and their corresponding descriptors are shown in \figref{reference_siti}. 
Furthermore, we report the spatial and temporal complexity \cite{menon2022vca,menon2023green} of the selected reference videos in \figref{reference_complexity}. As shown in both figures, the reference videos vary in spatial and temporal information, complexity, dynamic range, and median luminance, ensuring a balanced and diverse set of HDR content in the dataset.
This diversity allows for an accurate analysis of the effect of the viewing distance and display luminance on distortion visibility across a diverse set of content.



\begin{figure}[t]
\centering

\begin{subfigure}[t]{0.48\linewidth}
    \centering
    \includegraphics[width=\linewidth]{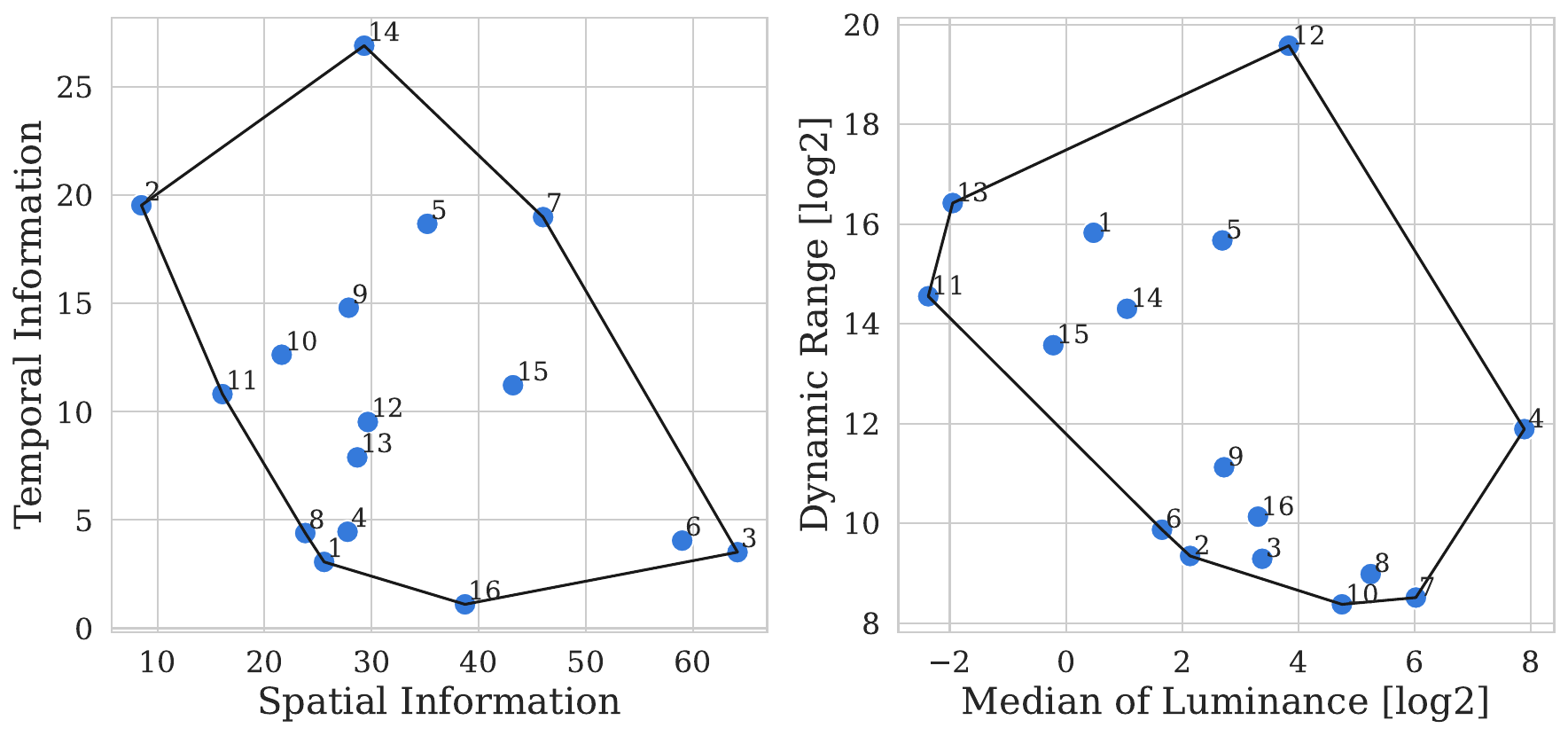}
    \caption{\edit{Scatter plots and corresponding convex hulls of the selected reference videos' descriptors.}}
    \label{fig:reference_siti}
\end{subfigure}
\hfill
\begin{subfigure}[t]{0.48\linewidth}
    \centering
    \includegraphics[width=\linewidth]{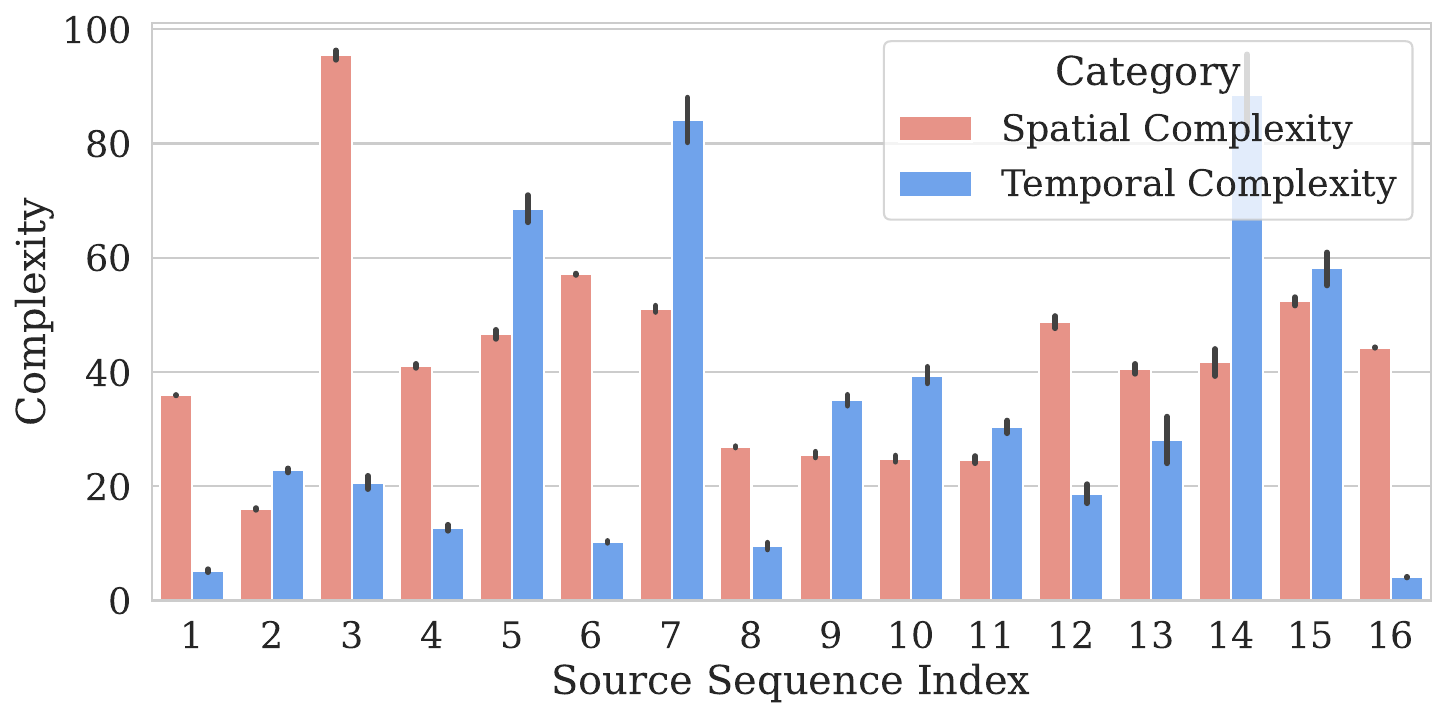}
    \caption{Spatial and temporal complexity of the selected reference videos.}
    \label{fig:reference_complexity}
\end{subfigure}

\caption{Characterization of the selected reference videos. 
Indices correspond to the video indices in \figref{reference_screenshots}. 
\edit{Differences between SI/TI and complexity measures arise from their distinct formulations: SI/TI operate on pixel-level gradients and frame differences, whereas complexity measures block-wise DCT texture energy.}}

\label{fig:reference_indicators}

\end{figure}


\subsection{Distortions}

Our focus in this study is to measure the effect of the display and viewing conditions on the perception of streaming distortions. Thus, it is necessary to design a concise distortion space that spans a broad range of quality degradations. 
The distortion space consists of two main distortions arising from video streaming: compression and upscaling artifacts. We employed the SVT-AV1 codec (v1.5.0) \cite{norkin2020svt,kossentini2020svt} to introduce the compression artifacts, as it is a representative of modern video codecs and offers good performance and encoding speed. To introduce upscaling distortions, we employed the Lanczos filter ($a=3$) \cite{duchon1979lanczos}, as it is a standard scaling filter used in many streaming pipelines. 

To obtain the distorted videos, each 4K source sequence was downscaled to the resolutions: 4K ($3840 \times 2160$ / unchanged), 1080p ($1920 \times 1080$), or 720p ($1280 \times 720$). 2K source sequences were either kept unchanged or downscaled to 720p. The videos were then encoded with the SVT-AV1 at three levels, producing nine test videos for each 4K source video and six for each 2K source video, totaling 132 test videos from the 16 reference videos. The encoding was done using a SVT-AV1 preset of four, as it is a common factor used in streaming to balance efficiency and compute time.



Because of the diversity of complexity of source sequences (as shown in \figref{reference_complexity}), it was necessary to select the CRF values for each source sequence individually to ensure a similar quality separation between the three compression levels.
To do so, each source sequence, upscaled to the display resolution of 4K, was compressed using a range of CRF values from 1 to 63. The ColorVideoVDP metric (v0.3) \cite{mantiuk2024colorvideovdp} was then used to guide the selection of the three optimal CRF values for the sequence, such as the distance in quality between the low and medium-quality videos is similar to the distance between the medium and high-quality videos (we used this condition under the assumption of the linearity of the JOD predictions of the metric). Moreover, we ensured that the high-quality videos achieved a perceptual quality exceeding 9.5 on a 10-point scale compared to the uncompressed video. For reproducibility, we report the selected CRF values for each source sequence and the command line we employed in \appref{video_encoding_parameters}. 
\revision{While some compression/upscaling distortion combination may not fully express a real-life streaming application, the choice was made to ensure a wide range of distortion levels.}


\edit{Our experiment design required simultaneous playback of three 4K HDR videos at 60 fps (two test videos and a reference video, as explained in the following section), which we could not read from uncompressed video files at the required frame rate using our hardware. We therefore used instead the highest-quality compressed version of each video, encoded at its native resolution using the highest-quality CRF setting (ranging from 5 to 11), as the reference condition. All selected reference videos achieved JOD scores above 9.5 (using ColorVideoVDP). Since a difference of 1 JOD (corresponding to a score of 9 JOD) marks approximately the onset of perceptible distortions, scores above 9.5 indicate that the compression was below the threshold of perceptual visibility. We further verified this through side-by-side visual inspection against the corresponding source videos and observed no visible compression artifacts. 
} 

\subsection{Display and viewing conditions}\label{sec:display_viewing_conditions}

We conducted the experimental study using two LG OLED Evo G2 55-inch 4K HDR10 displays, both operating on firmware v3.0.10. The displays were used to present the pair of videos being compared, with each test video shown on its respective display screen. The details on display configuration can be found in Section 1.2 of the supplementary. 

Following the BT.2100-2 recommendations \citep{itu2100hdr} for HDR viewing, we selected two viewing distances and two display luminance levels. The viewing distances considered were $1.6$ and $3.2$ times the display height, corresponding to $1.1$ and $2.2$ meters, and an effective resolution of 60\ppd and 120\ppd (pixels per visual degree). The two display luminance levels considered were a bright display, characterized by the original color-graded values as reproduced by the display (mean of 51\cdms, peak of 700\cdms), and a dimmed display, in which we reduced the luminance of the content by a factor of 8 in the linear RGB BT.2020 color space (mean of 5.6\cdms, peak of 600\cdms) before sending it to the display. \edit{The latter condition simulates dynamic dimming found in many TVs and handheld devices, which is intended to save power and improve comfort in dark environments. Note that this luminance manipulation is different from simulating displays of different peak luminance, e.g., HDR 400 and HDR 1000 displays. We found the factor of 8 to be the maximum that does not result in excessive loss of contrast, colour and visibility.} \edit{Because the original content contained higher luminance than the peak luminance of the OLED display (700\cdms{}), the dimming reduced the mean much more than the peak; brighter details could be reproduced in the dimmed condition without saturation.}
Unlike the work of \citet{ebenezer2024hdr}, we used the same display for both luminance conditions. This ensured that judgments were unaffected by display differences. Mikhailiuk et al. used an ND filter to dim the display \cite{mikhailiuk2021effect}; this approach was impractical due to the size of our displays \edit{and unnecessary as our OLED display could produce sufficiently low luminance levels.} 

\edit{The HDR content was displayed using the display's internal tone-mapping, with no external tone-mapping stage applied. \revision{Luminance values exceeding the display's capability were therefore processed by the display's internal tone-mapping algorithm rather than being clipped at the nominal peak luminance.} As illustrated in \figref{display_model_ramp_results}, the display response can be modeled for arbitrary input values, allowing us to visualize the resulting content and gamut for both display luminance conditions; these visualizations are provided in \figref{reference_videos_display_model}. This characterization also allows us to account for the effective dynamic range actually presented to observers under each condition. Details on how we measured the peak and mean luminance can be found in Section 1.3 of the supplementary.} 
The experiment was conducted in a dark environment.

\subsection{Experimental procedure}\label{sec:experimental_procedure}

The primary goal of this study is to assess the visibility of distortions as accurately as possible across display luminance levels and viewing distances. To achieve this, we adopted a pairwise comparison protocol \cite{perez2017practical}, as it relies on a simpler task suited for non-expert observers, and offers higher sensitivity compared to direct rating methodologies, such as the absolute category rating (ACR) \cite{Mantiuk_2012_cmp4met}. The sensitivity of the protocol is crucial when measuring the effect of the display and viewing conditions on the perceptual quality, which could be relatively subtle compared to the measurement noise. Additionally, to maximize the information gain from each comparison while concurrently minimizing the total number of required comparisons, we employed an active sampling method (ASAP \cite{mikhailiuk2021active}), which selects the best pairs to compare based on previous comparisons. \revision{The method constructs the minimum spanning tree in which the nodes are compared conditions and the edges are the comparisons. The tree is built so that each condition is compared to another one at least once and the comparisons maximize the information gain, measured as a KL-divergence before and after collecting new datapoints.}

\begin{figure}
\centering
\includegraphics[width=\columnwidth]{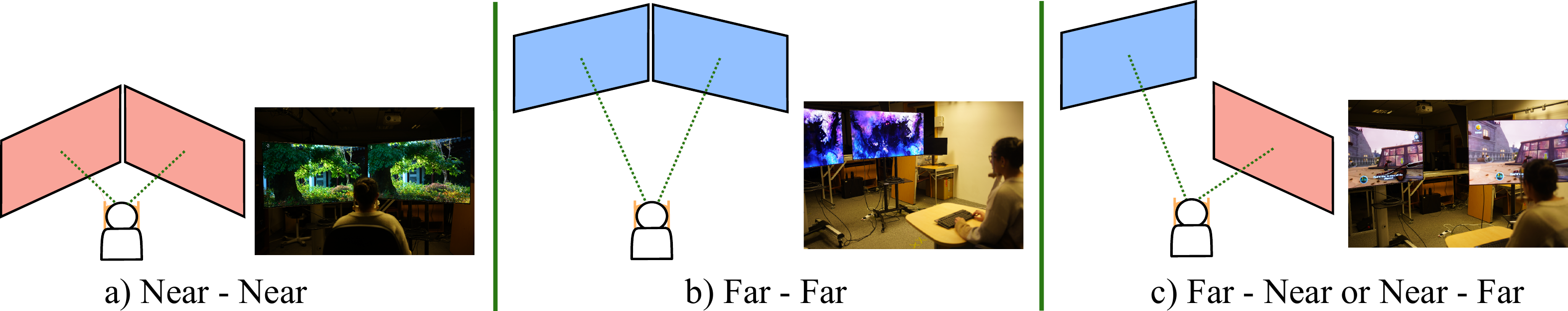}
\caption{The configurations corresponding to the combination of two viewing distances: Near-Near, Far-Far, Far-Near, and Near-Far (we show one of the two cross-viewing distance settings in this figure).}
\label{fig:viewing_settings}
\end{figure}

The experiment consisted of four sessions and a total of 960 trials. \revision{The four sessions corresponded to four display configurations: near-near, near-far, far-near, and far-far, where the names indicate the viewing distance of the left and right displays, respectively. Three of these configurations are shown in \figref{viewing_settings}. The near-far and far-near configurations allowed us to compare conditions across viewing distances. The screens were positioned such that the viewing distance between the chinrest, corresponding approximately to the participant's eye position, and the center of each display was maintained, and the viewing angle remained perpendicular to the center of each screen. When rotating the displays, we ensured that no specular reflections from one display to the other were observed by the participants and that the diffuse reflections were negligible.}

In each trial, observers were presented with a pair of test videos of the same content, simultaneously played on the two displays, with each screen showing one of the test videos. \edit{The assignment of the two test conditions to the left and right displays was randomized. For the near-near and far-far configurations, each condition was randomly assigned to one of the two displays. For comparisons across viewing distances, both the near-far and far-near configurations were used, thereby reversing the viewing distance between the left and right displays and reducing potential side- and display-specific biases}. Participants could also view the corresponding reference video by pressing and holding the space key; the reference video was then shown simultaneously on both displays. \edit{Thus, although the comparison involved two test videos, the experimental setup required simultaneous decoding of three 4K HDR video streams: the two test videos and the reference video.} To prevent participants from using temporal changes to identify the distortions, a black screen was presented for 0.5 seconds between transitions from the test to the reference video and vice versa.

Participants were instructed to select the test video that is closer and more similar in quality to the reference, by pressing the left or right key for the left or right display, respectively (a green box highlighting their choice was displayed to confirm their selection). The comparison with respect to the reference allowed us to assess the fidelity rather than the overall preference, a crucial consideration particularly when comparing the videos across viewing distances. To ensure that the instructions were carried out correctly, participants were required to view the reference video at least once per trial. The instructions were displayed on the screen after each trial. 

The participants had the flexibility to view both reference and test videos multiple times before making the selection, as the videos were presented in a loop, and the user could switch between the reference and test videos at any time. \revision{The unlimited presentation time was meant to improve the accuracy of the measurements.}

Each session was divided into two blocks, each featuring a different display luminance level: bright or dim display. We did not conduct cross-luminance comparisons due to the observers' bias toward the brighter display that we found in the preliminary pilot experiment results (which we did not find for cross-viewing distance comparisons). 
During the transition between the two blocks of the session, a random noise texture with mean luminance corresponding to the next luminance level was displayed for 20 seconds to adapt the eye to the new luminance level. The order of the sessions and blocks was randomized across participants to minimize any potential learning effect.




\subsection{Participants}

Thirty volunteers (16 females, 14 males), aged 13--52 years (mean 26, standard deviation 8), participated in the experiment. Twenty-five completed all four sessions, one participant completed one session, and four participated in all four sessions but did not complete all trials in one or more sessions. In total, we collected 25,762 pairwise comparisons (the number of comparisons between conditions is reported in Section 1.4 of the supplementary). 
All participants had normal color vision, confirmed using the Ishihara test, and normal or corrected-to-normal visual acuity, with a minimum acuity of 20/25 as determined by the Snellen test.

Before the experiment, participants received a briefing document and signed a consent form. Participants under 18 were accompanied by a legal guardian, and consent was provided by both. During their first session, participants also completed a brief training session to familiarize themselves with the setup and instructions. Sessions included 5--10 minute breaks after every 25 minutes, and participants could request additional breaks at any time using the `F' key. Participants completed at most two sessions consecutively, with a minimum 10-minute break between them and an hour-long break before a third session. In practice, all participants had approximately 24 hours between sessions. None reported significant fatigue affecting the experiment. The study was approved by the departmental ethics board, and participants were compensated.

\section{Experimental results}\label{sec:experimental_results}

\subsection{Subjective score scaling}
\label{sec:subjective_scaling}

The pairwise comparison responses from all observers were scaled (no outliers have been detected, as detailed in Section 2.1 of the supplementary) 
using Thurstone’s Case V model \cite{perez2017practical} into Just-Objectionable-Difference (JOD units) scores. The difference of 1\jod unit between the two conditions signifies that one condition has been selected as closer or more similar to the reference 75\% of the time compared to the other one. This approach allows us to derive quality scores comparable to other datasets employing the same JOD units.

Since pairwise comparisons were not conducted between conditions of different content or display luminance levels, the data was scaled separately for each content and display luminance level. 
All reference videos, irrespective of the viewing distance or display luminance level, were assumed to be the same node when scaling and then assigned a nominal quality of 10\jod (lower values referred to lower quality). 
\edit{Because JOD values represent the perceptual distance from the reference, using the same reference value provides a common anchor across the independently scaled conditions, allowing the resulting quality scores to be compared across content, viewing distances, and display luminance levels.}
Furthermore, confidence intervals of 95\% were obtained for each condition by bootstrapping with 5000 samples \cite{perez2017practical}.

Additionally, for analysis purposes, we scale the data across all contents (by scaling across all distortions and display and viewing conditions without taking into account the content) using the previously described scaling methodology. This allows us to analyze the effect of display and viewing conditions on each distortion independently of the content. These results are not directly included in the quality dataset; they are only utilized in our analysis, presented in the next section. 

\subsection{The effect of viewing distance and display luminance}\label{sec:effect_viewing_display}

\begin{figure*}
\centering
\includegraphics[width=\textwidth]{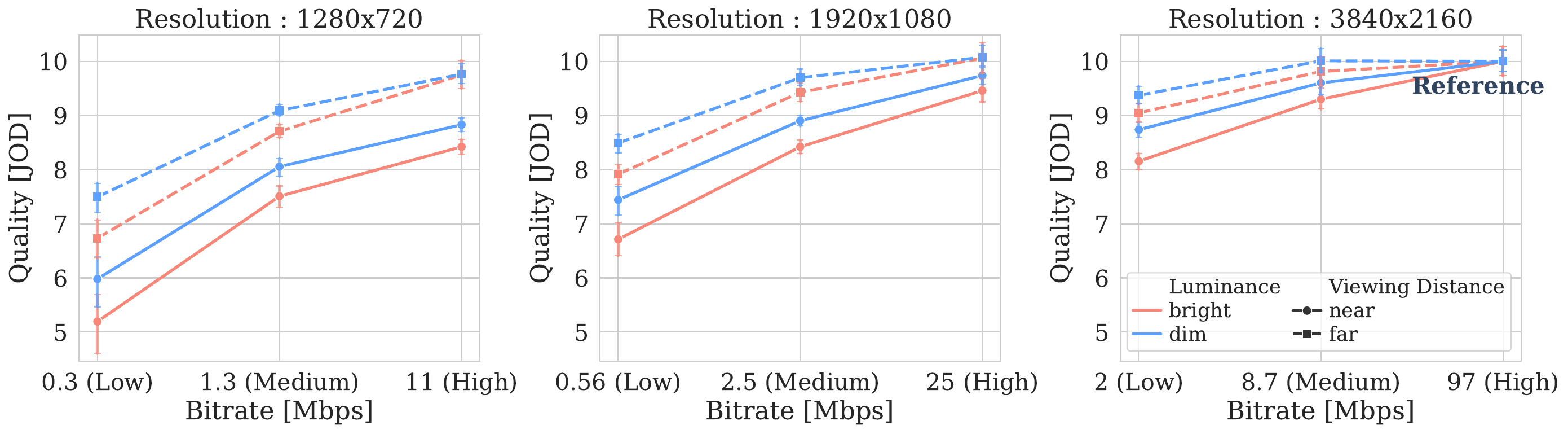}
\caption{The effect of viewing distance and display luminance level on the perception of quality for different resolutions and bitrates (compression levels) and across all contents. \revision{The bitrate reported in the figures is the average bitrate across all contents.} The error bars represent the 95\% confidence intervals.
}
\label{fig:effect_viewing_distance_peak_luminance}
\end{figure*}

To understand the effect of viewing distance and display luminance, we report the JOD scores across all participants and contents for each resolution, bit rate (compression level), display luminance level, and viewing distance in \figref{effect_viewing_distance_peak_luminance}.
From the figure, we can observe a consistent trend, wherein observers tend to perceive fewer distortions when videos are viewed from larger distances or on dimmer displays. This effect is more notable at lower resolutions and bit rates. For instance, at a low bitrate and a 720p or 1080p resolution, there is an approximate 2\jod difference in quality when comparing a video displayed on a dim screen viewed from a far distance with one displayed on a bright screen viewed from a near distance, meaning more than 91\% of observers are selecting the first condition when compared to the second one. 
Moreover, AV1 medium-level compression artifacts on 4K videos (no upscaling distortions) and upscaling distortions from 1080p to 4K (no compression artifacts) are imperceptible  ($\approx$10\jod) on dimmer displays seen from a far distance.

\begin{figure*}
\centering
\includegraphics[width=\textwidth]{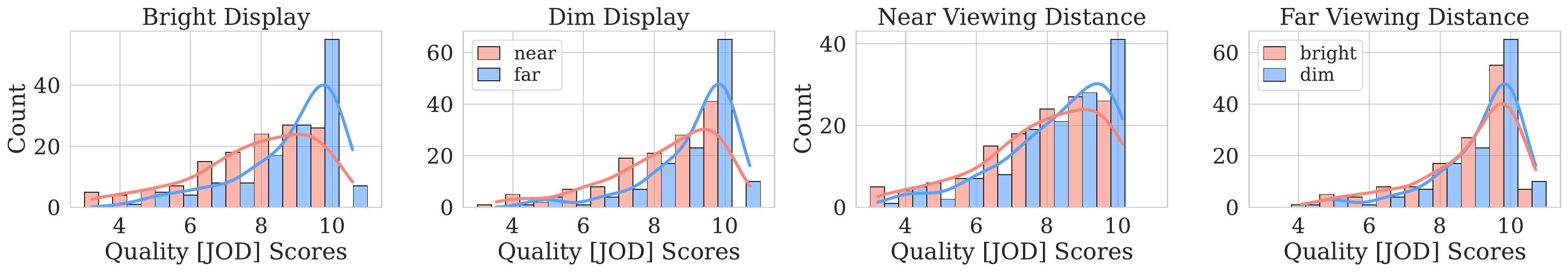}
 \caption{Histograms of the JOD scores and their distribution (estimated using the kernel density estimate for visualization purposes) across viewing distances and display luminance levels. 
 A few conditions exceed the score of 10\,JOD due to the measurement noise.}
\label{fig:histograms}
\end{figure*}

\begin{figure*}
\centering
\includegraphics[width=\textwidth]{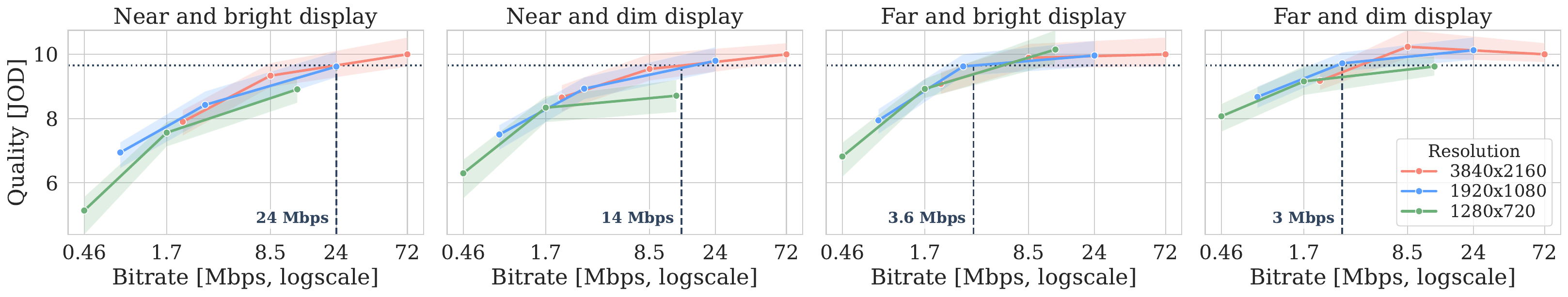}
\caption{Quality-bitrate curves for the RushPixar content (index 10 in \figref{reference_screenshots}) for each viewing distance and luminance level. The shaded areas represent 95\% confidence intervals. Dashed lines mark the approximate bitrate needed to reach 9.5\,JOD. A few conditions exceed the score of 10\,JOD due to measurement noise.}
\label{fig:bd_rate_RushPixar}
\end{figure*}


These observations clearly show that the effect of the viewing distance and display luminance level on the visibility of distortions is substantial. To test the statistical significance of these effects, we employ a 2-way analysis of variance (ANOVA). The results showed a significant effect of both viewing distance ($F(1,13917)=1598.6$, $p<0.001$) and display luminance level ($F(1,13917)=238.5$, $p<0.001$) on the JOD scores. The effect size of the viewing distance was 0.101, indicative of a medium effect on distortion perception, while the effect size for the display luminance level was 0.015, reflecting a small effect on the perceptual quality. 

\edit{The effect of viewing distance was largely independent of the content characteristics considered (spatial, temporal information, dynamic range, median luminance, and spatial, temporal complexity), whereas the effect of display luminance level was more pronounced for darker contents, showing a significant correlation with median luminance ($r=-0.55$, $p=0.027$); further analysis of these content dependencies is provided in \appref{effect_size_anova_vs_content}}. 

This significant effect is further confirmed when analyzing the distribution of the quality scores of test videos for each viewing distance and display luminance level, as illustrated in \figref{histograms}. The two leftmost columns show the effect of the viewing distance on the distribution of the JOD scores at each fixed luminance level. Conversely, the two rightmost columns illustrate the effect of the luminance level on the distribution of the JOD scores at each fixed viewing distance. These curves illustrate once again an evident increase in the perceptual quality of videos for a lower luminance level and/or larger viewing distance, reaffirming that distortions are less perceptible for such display and viewing conditions.
Moreover, we can observe from the figure that the effect of viewing distance decreases as display luminance decreases. Conversely, the effect of display luminance level decreases as viewing distance increases. These findings emphasize the interdependent nature of these conditions in their impact on the visibility of distortions, thereby showcasing the importance of accounting for all display and viewing conditions when assessing video quality.


To investigate the impact of these conditions on video streaming optimization, we plot the quality-bitrate curve, often referred to as the bitrate ladder, for the RushPixar content across the combinations of the two viewing distances and display luminance levels in \figref{bd_rate_RushPixar}.
The plots validate the overall trend, showcasing the significant effect of the viewing distance and display luminance on the visibility of distortions in videos, in which the perceptual quality is higher for a dimmer display viewed from a far viewing distance.
These effects can be employed to further optimize video streaming. For instance, consider the scenario where a user requests the RushPixar content with a high quality of 9.5\jod. Assuming the optimal user conditions (near viewing distance and bright display), a bit rate of 24\mbps would be required to stream the video. However, when accounting for the user conditions, for example, a larger viewing distance and a dimmer display, the same quality can be achieved with only a bit rate of 3\mbps, representing an eightfold decrease in bitrate requirements for streaming. 

\section{Quality Metric Evaluation}\label{sec:quality_metrics_evaluation}



\begin{figure}
\centering
\includegraphics[width=\columnwidth]{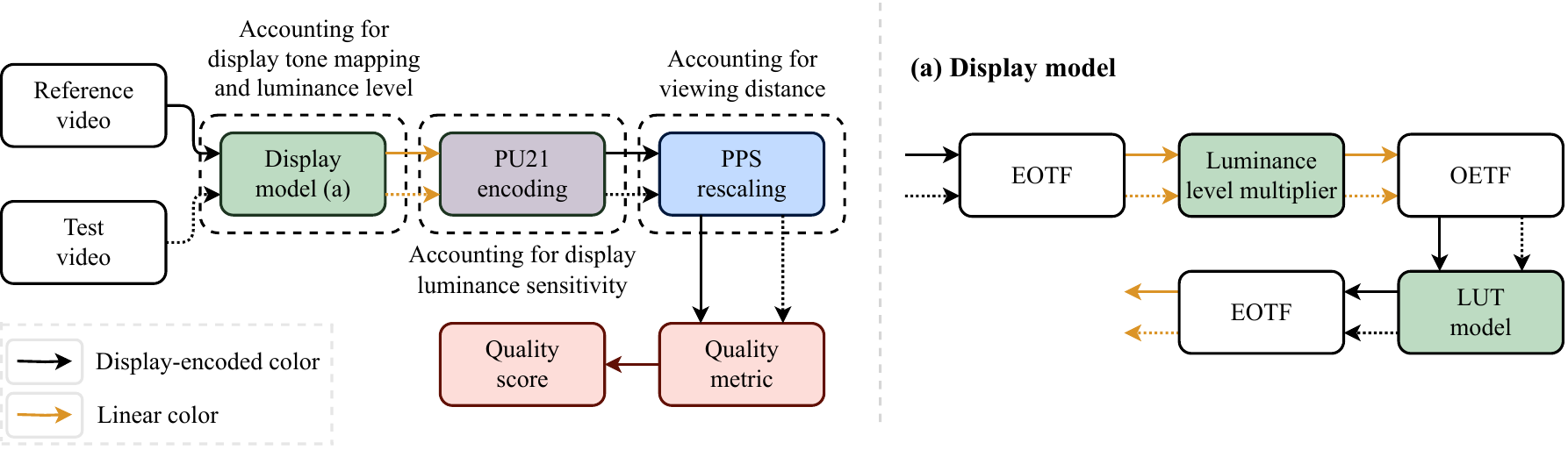}
\caption{
\revision{An illustration of the pipeline used to simulate the human visual system perception of HDR videos across viewing distances and display luminance levels, prior to the metrics evaluation. The display model component is employed for all metrics. However, metrics that natively account for the viewing distance do not employ PPS rescaling, and photometric metrics that utilize linear RGB input do not use the PU21 encoding.}}
\label{fig:metrics_pre_processing}
\end{figure}

In this section, we evaluate how well existing image and video quality metrics predict the quality of the HDR-VDC dataset, specifically their performance in predicting the effects of viewing distance and display luminance. Because many of the metrics do not account for these factors, we introduce pre-processing steps that model human visual perception of HDR videos at different viewing distances and luminance levels. We use these pre-processed videos to adapt the standard quality metrics for these conditions. An illustration of the processing pipeline is shown in \figref{metrics_pre_processing} and detailed in the following sections.

\subsection{Display model and tone mapping}\label{sec:lut_model}


Because the color gamut and luminance range found in HDR content (BT.2020 color representation) can much exceed the capabilities of an HDR display, HDR displays typically use tone mapping algorithms to accommodate content to their color capabilities. To account for such tone mapping, we measured display response and fitted a forward display model for the LG OLED G2 display, following \cite{hung1993colorimetric, ashraf2024color}. \edit{This characterization was not intended to compensate for display limitations, but to account for the display as part of the end-to-end video presentation pipeline. This allowed us to pass to the quality metrics the colors and light levels actually presented on the screen, rather than those stored in the video files}. 

The color measurements were collected by displaying uniform color patches centered on the screen, covering 5\% of the screen area. Two measuring instruments were used: the Konica Minolta Chroma Meter CS-200 and the JETI Specbos 1211 broadband spectroradiometer. The JETI Specbos was used to measure the light emitted from the display for input linear RGB values exceeding or equal to 1\cdms as it provided faster measurement. The Konica Minolta was used for values below or equal to 1\cdms due to its higher sensitivity at lower light levels. For input linear RGB values at exactly 1\cdms, measurements from both equipment were averaged to ensure a smooth transition between values measured with different devices. 
We measured a 10$\times$10$\times$10 grid of linear RGB BT.2020 values ranging from 0.015 to 10\,000\cdms spaced logarithmically to cover the entire display gamut. Additionally, we set the background luminance of the screen (outside the 5\% color patch) to the median luminance of all reference videos. This allowed us to account for the effect of neighboring pixels’ luminance on the measured light from each pixel.
The measurements were encoded using the PQ function and then stored as a look-up table (LUT). We employed the PQ encoding as it resulted in higher accuracy. 

To evaluate the model's performance on unseen colors, we measured 200 random colors not included in the LUT using the Konica Minolta CS-200 and compared them to the predictions of the LUT model (we employed linear interpolation to predict the colors using the LUT). The accuracy of our model was assessed using the $\Delta E_{00}$ color difference metric~\citep{sharma2005ciede2000}. The results showed a mean $\Delta E_{00}$ of 3.07 and a median $\Delta E_{00}$ of 2.41.
Although these values might initially appear large for color errors, they represent a relatively small error in the context of measuring OLED displays \cite{ashraf2024color}.

\begin{figure}
\centering

\begin{subfigure}[t]{0.48\columnwidth}
    \centering
    \includegraphics[width=\linewidth]{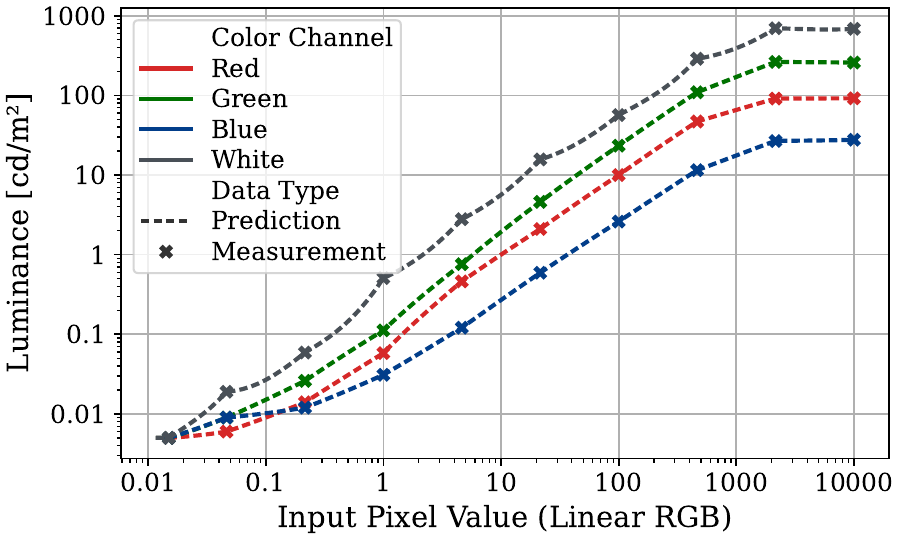}
    \caption{\revision{The display response of the red, green, blue, and white color channels, as measured from the display (marker points) and predicted using our LUT model. The small curvatures in the white channel are due to the interpolation in the PQ-space.}}
    \label{fig:display_model_ramp_results}
\end{subfigure}
\hfill
\begin{subfigure}[t]{0.48\columnwidth}
    \centering
    \includegraphics[width=\linewidth]{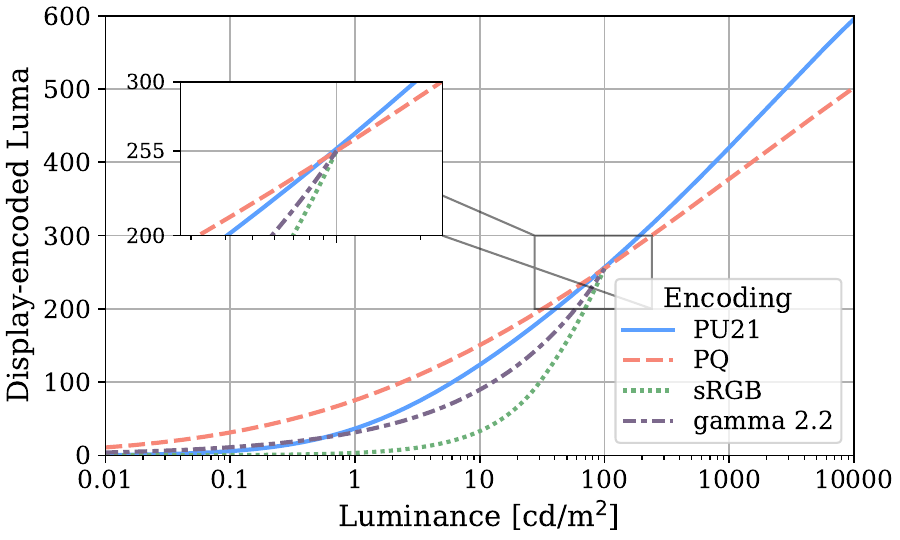}
    \caption{Comparison of the PU21, PQ, sRGB, and gamma ($\alpha=2.2$) encoding functions. Luminance is shown on a logarithmic scale, while luma is shown on a linear scale. All functions are rescaled such that 100\cdms{} maps to 255, corresponding to the typical SDR range.}
    \label{fig:pu21_comparison}
\end{subfigure}

\caption{(a) Display model response and (b) comparison of encoding functions.}
\label{fig:combined_results}

\end{figure}

Using this look-up table, we defined our display model, that accounts for both the display tone mapping algorithm, and the display luminance level, as shown in \figref{metrics_pre_processing}-(a).
The difference in metrics performance when employing the look-up table (excluding the part where we account for the display luminance level) was not statistically significant (refer to Section 3.2 of the supplementary 
for an in-depth analysis), \revision{which is likely due to the fact that the tone mapping employed in OLED displays is mostly linear up to 1000\cdms{} (as observed in \figref{display_model_ramp_results}); therefore it affected only a small portion of the brightest pixels.} 
However, we assert that the LUT display model is essential to faithfully represent content as seen by the observers. Similarly, metrics should be tested and calibrated on such observed content rather than encoded data. 


\subsection{Accounting for luminance sensitivity}

The visibility of distortions varies significantly depending on the absolute luminance level (in nits or \cdms{}) at which the content is presented. In particular, distortions become less visible at lower luminance levels, at which the visual system is less sensitive \cite{ashraf2024castlecsf}. 
To account for this effect, Aydın \etal{} \cite{aydin2008extending} proposed a perceptually uniform (PU) encoding derived from contrast sensitivity models. \revision{Ebenezer \etal{} \cite{ebenezer2023making} proposed a local expansive nonlinearity mapping ``HDRMAX'' which enhances the visibility of distortions at the brightest and darkest local regions of the content. We used a modernized version of the PU encoding, known as PU21, which incorporates modern contrast sensitivity data, accounts for display glare, and is based on a model for detecting banding artifacts \cite{Mantiuk_Azimi_2021}. We incorporated the PU21 encoding, because it does not necessitate re-training the metrics and show higher performance. Nonetheless, we included the metrics pre-trained with the HDRMAX features in \secref{metric_evaluation_protocol}.} An illustration of the PU21 encoding in comparison to standard HDR encoding functions (Perceptual Quantizer PQ) and SDR encoding functions (sRGB and gamma) is shown in \figref{pu21_comparison}.

As shown in \figref{metrics_pre_processing}, the PU21 function is applied to linear RGB values to encode those into perceptually uniform units that are passed to the quality metric. Given that the PU21 output scale is relative, we scaled the output so that 700\cdms{} in linear RGB values (the maximum luminance of the LG OLED G2 display used in the experiment) corresponds to 1 in perceptually uniform units. This ensures that the input frames fall within the [0,1] range, which is accepted by the quality metrics.
Because PSNR and SSIM can operate on the arbitrary color ranges, we scaled the output for these two metrics so that 100\cdms{} in linear RGB values (the peak luminance of the displays for which these metrics were designed) corresponds to 1 in perceptually uniform units. This ensures that the quality values (e.g., PSNR) for HDR are comparable to those reported for SDR content.



\subsection{Accounting for the viewing distance}

Metrics that do not account for viewing distance can be adapted by rescaling the input images and videos \cite{gu2013self, hammou2023comparison, hammou2024hvei}. This method is based on the observation that the resolution of the image projected on the retina is proportional to the viewing distance. As the viewing distance increases, the retinal image becomes smaller and the perceived image shifts towards higher spatial frequencies.

\revision{Multiple rescaling factors have been proposed previously \cite{gu2013self, hammou2023comparison, hammou2024hvei} to simulate the effect of viewing distance on the perception of images or videos. However, we found that the commonly used self-adaptive scale transform (SAST) \cite{gu2013self} to be unsuitable as it results in different scaling factors depending on the video aspect ratio. The visibility of distortions is unlikely to be affected by this. Instead, we employ perspective projection scaling (PPS), which yielded the best results across most compared metrics. 
For a more detailed description of each approach and the results of their comparison, refer to Section 3.3 of the supplementary. 
}


\revision{PPS is based on the assumption that metrics are calibrated to perform at an optimal effective resolution $\rho_m$ in pixels per degree. Knowing the stimulus effective resolution $\rho$, the rescaling factor $z_{\text{PPS}}$ is given thusly:
\begin{equation}
    \label{eq:PPT}
    z_{\text{PPS}} = \frac{\rho_m}{\rho}
\end{equation}
with
\begin{equation}
    \label{eq:PPD}
    \rho = \frac{\pi r_{y}}{360 \text{atan}\left ( \frac{0.5 h}{d} \right)}
\end{equation}
where $r_y$ is the display's vertical resolution in pixels, $h$ is the display height in meters or inches, and $d$ is the viewing distance in the same units as $h$. The metrics' effective resolution $\rho_m$ is empirically defined as 60 pixels per visual degree (ppd).}

\revision{By rounding the rescaling factor to the first significant digit, we found that $z_{\text{PPS}} = 1$ for a near viewing distance ($D = 1.6 H$) and $z_{\text{PPS}} = 0.5$ for a far viewing distance ($D = 3.2 H$). The rescaling has been applied to PU21-encoded frames, as shown in \figref{metrics_pre_processing}, using the bicubic filter as it has been shown to perform well in this task \cite{hammou2024hvei}.}


\begin{figure*}
\centering
\includegraphics[width=\textwidth]{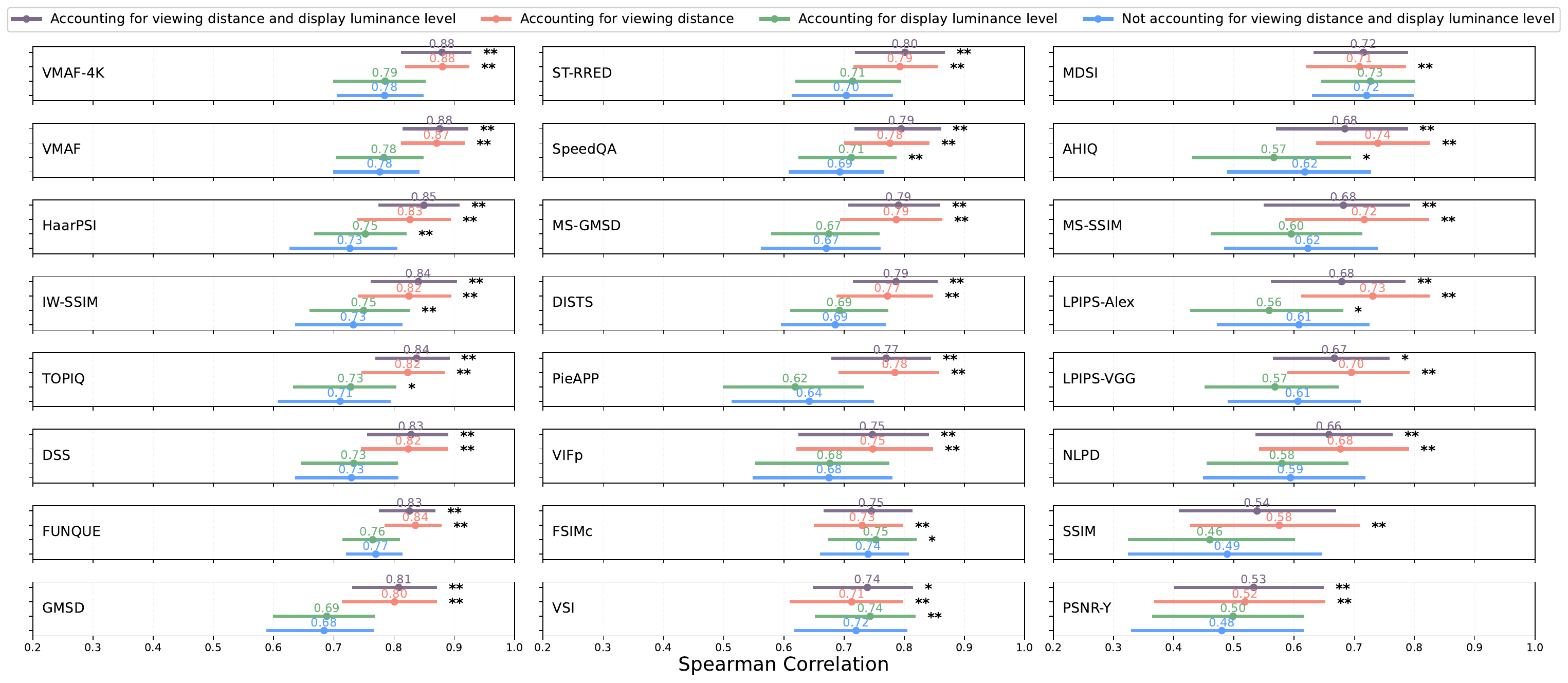}
\caption{\revision{SROCC performance of the metrics when accounting for the viewing distance and/or the display luminance level (by employing the PPS rescaling and the luminance level multiplier presented in \figref{metrics_pre_processing}) and when not accounting for the viewing distance and display luminance (no PPS rescaling and luminance level multiplier). Error bars denote 95\% confidence intervals. The * or ** symbols indicate a statistically significant difference in performance from the baseline (not accounting for the viewing conditions) at $p < 0.05$ and $p < 0.01$ (one-tailed paired t-test), respectively.}}
\label{fig:effect_vd_luminance_comparison_metrics_performance}
\end{figure*}

\subsection{Metric Evaluation Protocol}
\label{sec:metric_evaluation_protocol}


We selected 35 state-of-the-art full-reference image and video quality metrics and evaluated their performance on the HDR-VDC dataset. We selected full-reference metrics, as we only measured the visibility of distortions in the dataset. The details of the metrics are provided in Section 3.1 of the supplementary. 
We do not re-calibrate any of the metrics on the dataset, as we want to evaluate how well the metrics generalize to this dataset, which is why we selected metrics that are widely used on streaming distortions. \edit{Moreover, we used the reference videos selected for the experiment as the reference inputs to the metrics, rather than the uncompressed source videos. This ensures that the metric evaluation matches the reference condition used to collect the subjective scores and therefore evaluates the same distortions presented to the observers}.
As shown in \figref{metrics_pre_processing}, the metrics were adapted to account for the effect of viewing distance, display luminance level, and HDR content. Photometric metrics did not require the PU21 encoding, and metrics that natively account for the viewing distance did not require the PPS rescaling. Additionally, three metrics --- SSIM, MS-SSIM, and VMAF --- were adapted and re-calibrated to account for luminance sensitivity using the HDRMAX feature set \cite{ebenezer2023making}, as provided by the authors.

A common weakness of quality metric benchmark studies is that the results reported on one dataset often do not generalize to other datasets. To address this issue, we employed statistical testing that accounts for both main sources of uncertainty: (a) the dataset contains a limited sample of all possible content, and (b) the subjective quality scores have been measured for a limited number of observers (measurement noise). 
While many studies address the first uncertainty, few address the second. To robustly account for both factors, we generated a 5000-sample distribution of JOD scores by bootstrapping. This was achieved by randomly sampling with replacement both across the experiment participants and the video content (independently). We used those distributions to confirm that one metric performed better than another. 
We evaluated the metrics’ performance using five standard indices: Spearman Rank-Order Correlation Coefficient (SROCC), Pearson Linear Correlation Coefficient (PLCC), Kendall Rank-Order Correlation Coefficient (KROCC), Root Mean Squared Error (RMSE), and reduced Chi-Square ($\mathcal{X}_{v}^2$). Before running PLCC, RMSE, and reduced Chi-Square, we fitted the metrics’ predictions to JOD scores using a four-parameter logistic function, as described in \cite{antkowiak2000final}.  For PLCC, RMSE, and reduced Chi-Square, the logistic function parameters were optimized once for the entire bootstrap sample; they were fitted to the JOD scores for all images and observers without resampling.

To test whether we have statistical evidence for performance difference, we used a one-tailed paired t-test at $\alpha=0.05$ significance level. The test accounts for the covariance (dependence) between the distributions (due to bootstrapping). 
The Fisher Z-transform was used for the correlation indices distribution before the t-test to ensure their normality. The degree of freedom was set as $N - 1$, where $N$ represents the number of observers multiplied by the number of contents, $N = 480$ (for 30 observers and 16 contents). 
Finally, to ensure the significance level while making multiple comparisons, we applied the Bonferroni-Holm correction \cite{hothorn2008simultaneous} to the p-values.



\begin{figure*}
\centering
\includegraphics[width=\textwidth]{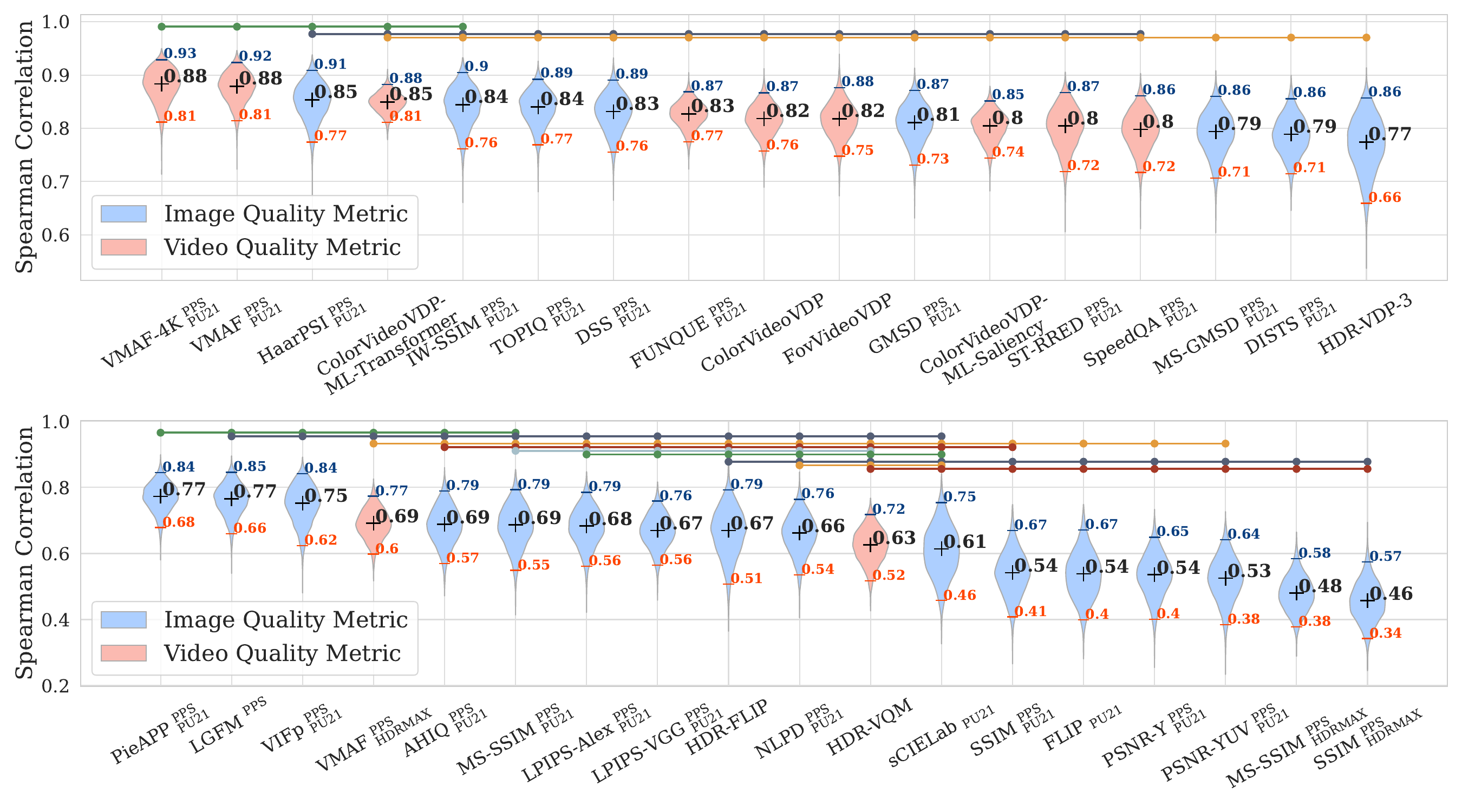}
\caption{SROCC performance indices of the metrics on the HDR-VDC dataset. The metrics are ordered from the best (top-left) to the worst (bottom-right).
Non-photometric quality metrics and/or metrics that do not natively account for viewing distance were adapted using PU21 encoding and/or PPS rescaling (as indicated in the label).  
The violin plots illustrate the distribution of the means that model variability due to content selection and observer responses. The black “+” denotes the mean score, and the red and blue “-” denote the 2.5th and 97.5th percentiles, which estimate the 95\% confidence interval. The solid lines group the metrics for which there is no statistical significance in the difference in their performance.
}
\label{fig:metrics_result_spearman_rmse}
\end{figure*}

\subsection{Importance of viewing conditions}

\revision{First, we test whether our proposed extension of the metrics, with PPS rescaling, and luminance level multiplier presented in \figref{metrics_pre_processing}, improves the performance of quality metrics that do not account for the display and viewing conditions. 
\figref{effect_vd_luminance_comparison_metrics_performance} illustrates the SROCC performance of the metrics under different modeling conditions. The blue bars represent the baseline, in which neither viewing distance nor display luminance is accounted for. 
The green, pink, and violet bars show the performance achieved when accounting for display luminance, viewing distance, or both factors simultaneously, respectively. }
\revision{Overall, most metrics exhibit statistically significant improvements when the effect of viewing distance is taken into account. The only exceptions are FSIMc, VSI, and MDSI, which could not be extended to incorporate viewing-distance effects because they internally rescale input frames to a fixed resolution of $480 \times 270$ before quality prediction. As a result, these metrics are excluded from further analysis. Conversely, most metrics do not show a statistically significant improvement over the baseline when display luminance is taken into account. This outcome aligns with our earlier finding that the overall impact of luminance level was comparatively small. We attribute this behavior to the fact that, although reduced luminance attenuated distortions in darker regions, distortions remained visible in brighter regions due to the relatively high peak luminance (600 \cdms{}) in the dim condition --- a consequence of display saturation at higher luminance levels (refer to \figref{display_model_ramp_results}). Moreover, two metrics, AHIQ, and LPIPS-Alex show statistically significant deterioration in performance when accounting for the effect of the luminance level. We believe that this is due to the modeling of the metrics, as both are deep-learning-based; we could not find why they could not be adapted to account for this effect. Nonetheless, most metrics perform significantly better when employing our pre-processing pipeline, which shows the benefit of adapting (selected) quality metrics to predict the effect of the display and viewing environment, especially in applications such as video streaming.}

\subsection{Metric results} \label{sec:metric_results}


We report the SROCC distributions for all the considered metrics in \figref{metrics_result_spearman_rmse}. The color lines above the distributions illustrate the results of paired statistical tests. If the metrics are connected by one of the lines, there is no statistical evidence that one metric performs better than the other. 


\figref{metrics_result_spearman_rmse} shows that the best-performing metrics are VMAF (both variants), HaarPSI, ColorVideoVDP-ML-Transformer, and IW-SSIM. The differences between each pair of metrics in this group are not statistically significant. Other metrics, such as TOPIQ, DSS, FUNQUE, ColorVideoVDP, FovVideoVDP, GMSD, ColorVideoVDP-ML-Saliency, ST-RRED, and SpeedQA, show statistically comparable results to the best-performing metrics, with the exclusion of VMAF. 
To maintain statistical rigor and ensure that our conclusions will generalize, we do not pick the ``winner'' metric. We note, however, that this group includes both the metrics that natively account for the viewing conditions (ColorVideoVDP and its variants) and those that rely on our pre-processing framework (\figref{metrics_pre_processing}), such as VMAF. \edit{Although ColorVideoVDP was used to select the CRF values, its role was limited to approximately spacing the quality levels, and the results provide no evidence of a substantial bias toward ColorVideoVDP or related metrics; nonetheless, a small residual effect cannot be excluded.}

These overall findings provide further evidence showing that many of the existing metrics can be adapted to account for the viewing conditions. 
Interestingly, the metrics that are widely used for evaluation, including PSNR (Y and YUV variants), sCIELab, SSIM, and LPIPS, performed very poorly in our tests. The results for the other performance indicators (PLCC, KROCC, RMSE, and reduced Chi-Square) and for each viewing conditions are included in Section 3.4 of the supplementary. 

\begin{figure}[!hbt]
\centering
\includegraphics[width=\columnwidth]{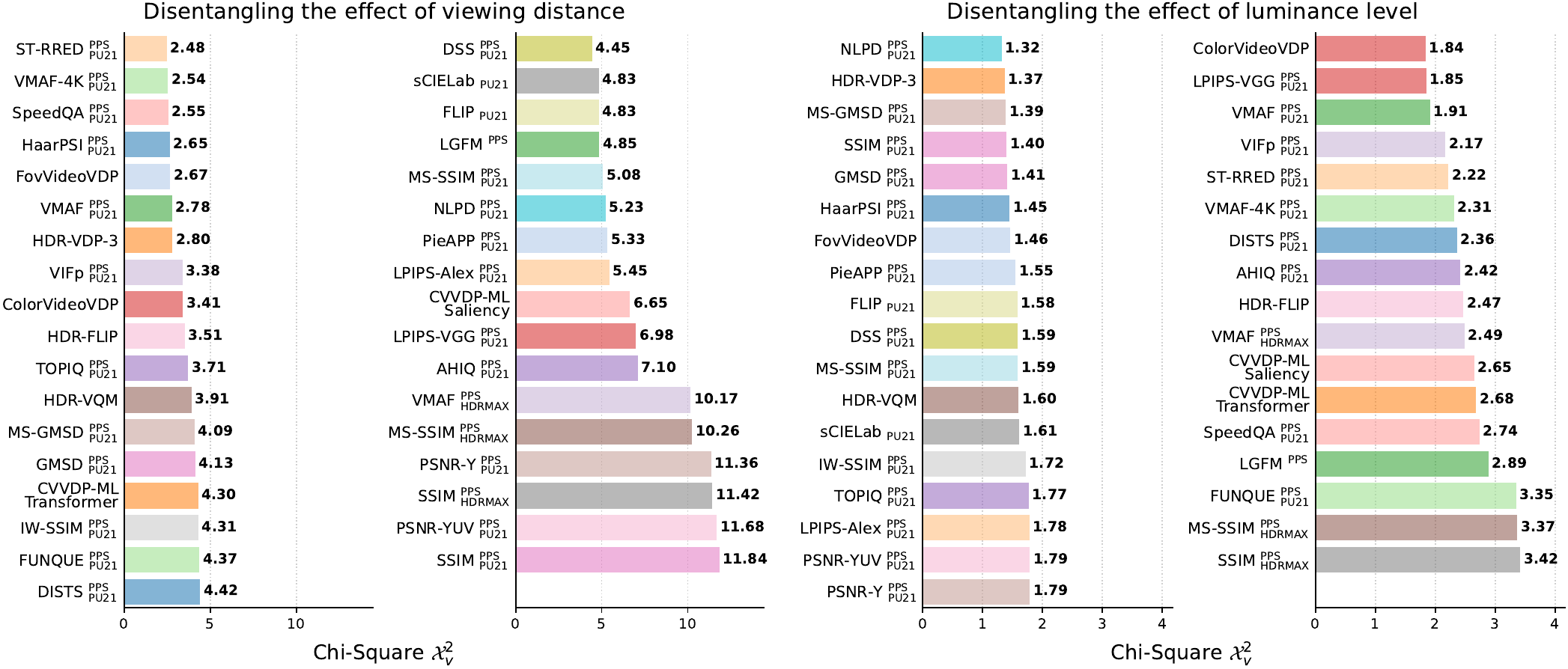}
\caption{\revision{The Reduced Chi-Square $\mathcal{X}^2_{v}$ performance (the lower, the better) of the disentanglement analysis of both the effect of viewing distance and the display luminance level.}}
\label{fig:disentangled_analysis}
\end{figure}

\subsection{The effect of viewing distance and display luminance disentangled}




In this paper, we aim to evaluate the effectiveness of the metrics in predicting the effect of viewing distance and display luminance on the visibility of distortions. The results shown in \figref{metrics_result_spearman_rmse} do not tell us whether the metrics with the highest SROCC are those that well account for the viewing conditions or these are just good metrics in general (can well evaluate distortions). Therefore, in this section, we aim to disentangle the metrics’ ability to assess the effect of viewing conditions independently from their general performance. 

To disentangle the effect of viewing distance (or display luminance level), we calculate the difference between the subjective scores at two viewing distances (or display luminance levels) as:
\begin{equation}
\Delta \text{jod}_{i,\text{vd}} = \text{jod}_{i,\text{far}} - \text{jod}_{i,\text{near}}\,,    
\end{equation}
where $\text{jod}_{i,far}$ is the JOD score for test condition $i$ at a far viewing distance and $\text{jod}_{i,near}$ is the JOD score for test condition $i$ at a near viewing distance. Similarly, we disentangle the effect of viewing distance (or display luminance level) from the metrics predictions by taking the difference between the pair of metric predictions. This methodology allows us to isolate the change in quality due to the change in viewing distance (or display luminance level) and evaluate how well a metric is capable of predicting this change. For metrics that predict quality values that are non-linearly related to JODs (or MOS), such as ST-RRED or SpeedQA, we first apply a logarithmic function to linearize the predictions before disentangling them.
To align the differences in metric predictions with the differences in JODs, we use a simple mapping function $f(x) = a \cdot x$, instead of a logistic function. This is because computing the difference effectively eliminates the non-linearity. We show the scatter plots of disentangled predictions in \appref{scatter_plots_figs}. 


In this analysis, we report reduced Chi-Square $\mathcal{X}_{v}^2$ between the disentangled JOD scores and the mapped disentangled metrics predictions. 
The reduced Chi-Square measures the goodness of the fit between the mapped disentangled metrics predictions and the disentangled JOD scores, indicating how well the metrics predict the changes in quality that are due to the change in the viewing condition (either viewing distance or display luminance level). We selected it because it accounts for the variance (noise error) in the JOD scores and can better illustrate the effect size. 
\revision{A $\mathcal{X}_{v}^2 = 1$ represents a good fit, which means the metric is capable of predicting the effect of the display or viewing condition effectively. A $\mathcal{X}_{v}^2 \gg 1$ represents a poor fit, which indicates that the metric is not capable of predicting the effect of the viewing distance or display luminance level. We report the results in \figref{disentangled_analysis}.}

\paragraph{Effect of viewing distance} \revision{ As \figref{disentangled_analysis} shows, the metrics that include CSF (HDR-VDP-3, HDR-FLIP, FovVideoVDP, ColorVideoVDP) show errors ranging from 2.67 to 3.51, which, on average, are lower than for metrics that need to rely on rescaling. However, several metrics perform well when combined with rescaling: ST-RRED, VMAF-4K, SpeedQA, HaarPSI, VMAF and VIFp.
Note that DSS, FUNQUE, and IW-SSIM performed very well on our dataset (refer to \figref{metrics_result_spearman_rmse}). Yet, our disentangled results show that those metrics do not account well for the effect of viewing distance.
Our results could be used to select metrics that are likely to perform well with rescaling and those that cannot be easily extended to account for the viewing distance.
}


\paragraph{Effect of luminance} \revision{Compared to the effect of the viewing distance, the effect of the luminance results in low errors across most metrics, most likely because the luminance has a smaller effect on quality than the viewing distance (see \figref{effect_vd_luminance_comparison_metrics_performance}). The metrics that natively account for the photometry of the display (FovVideoVDP, ColorVideoVDP, HDR-VDP-3, HDR-FLIP) perform well, with the error ranging from 1.37 to 2.47. However, it is interesting to note that HDR-FLIP performs worse than standard FLIP combined with PU21, this can be attributed to the use of multiple exposures in HDR-FLIP, which do not allow the metric to adapt well to a specific luminance level. Moreover, the metrics that rely on the HDRMAX features from \cite{ebenezer2023making} performed worse in our tests in comparison to PU21. This can be attributed to the local mapping employed in \cite{ebenezer2023making}, which could have introduced new artifacts or implified existing ones.} 

\section{Discussion}
\label{sec:discussion}

\edit{
Our results show that viewing conditions can substantially affect the visibility of streaming distortions and should therefore be considered when optimizing HDR video delivery. The results also show that quality metrics can account for these effects either through native modeling of the viewing conditions or, for some metrics, through appropriate pre-processing. We discuss the implications and limitations of these findings below.
}

\subsection{Implications for viewing-condition-aware streaming}
\label{sec:disc-streaming}

\edit{
The results presented in \secref{effect_viewing_display} have important implications for adaptive video streaming. The bitrate and resolution required to achieve a given perceptual quality depend strongly on the viewing conditions, with average bitrate savings of 78.1\% $\pm$ 10.4\% at 9~JOD and 85.5\% $\pm$ 7.3\% at 9.5~JOD across the 16 contents between the most demanding and least demanding viewing conditions. This suggests that streaming systems could adapt their bitrate and resolution ladders to the viewer's estimated viewing conditions, selecting the lowest-cost representation that achieves the target quality. The potential benefit is particularly relevant at lower bitrates and resolutions, where distortions are more visible. Practical deployment would require estimating viewing conditions on the client, for example from device characteristics, display settings, ambient-light sensors, or distance estimation. We do not evaluate such estimation or end-to-end condition-aware bitrate selection here, which we leave for future work.
}

\subsection{Implications for quality-metric development}
\label{sec:disc-metrics}

\edit{
The metric evaluation provides four main observations. First, the results indicate that our pre-processing adaptation performs as well as the native handling of physical luminance and viewing distance in some metrics (ColorVideoVDP, FLIP, HDR-VDP). However, our results are limited to two viewing distances and to the mid- and high-frequency streaming distortions represented in our dataset. We therefore do not expect the proposed adaptation to generalize reliably to shorter viewing distances, where the image must be enlarged, or low-frequency distortions (e.g., display vignetting, global color shifts). This is consistent with our previous findings~\cite{hammou2024hvei}, which showed that the proposed rescaling approximates the CSF primarily at mid and high spatial frequencies. Native modeling of viewing conditions within quality metrics therefore remains important for broader generalization.
Second, overall prediction accuracy and sensitivity to viewing conditions are distinct properties. Metrics that perform well in overall quality prediction do not necessarily reproduce the effects of changes in viewing conditions. This suggests that future benchmarks should evaluate not only overall prediction accuracy but also whether metrics correctly capture quality changes caused by viewing conditions.
Third, the effect of display luminance was more difficult to model. Most metrics did not show a significant improvement when luminance was accounted for, while AHIQ and LPIPS-Alex performed worse. This may partly reflect the relatively small luminance effect observed in our dataset and the relatively high peak luminance retained in the dim condition. Datasets covering a wider range of luminance levels are therefore needed to establish how reliably existing metrics can model luminance effects.
Finally, widely used metrics such as PSNR, SSIM, and LPIPS performed poorly in our experiments, highlighting the need for quality metrics that explicitly account for HDR display and viewing conditions.
}


\subsection{Generalizability and limitations}
\label{sec:disc-generalizability}

\edit{
The generalizability of our findings is limited by the range of conditions investigated. Viewing distance produced a medium effect ($\eta^2=0.101$), comparable to that reported for the SDR VLIC image dataset~\cite{mikhailiuk2021effect} ($\eta^2=0.109$; \appref{comparison_existing_datasets}), whereas display luminance produced a smaller effect ($\eta^2=0.015$). The luminance effect may become more pronounced under darker conditions or on displays with lower peak luminance, as suggested by the comparison with VLIC. More broadly, our results are most directly applicable to large HDR displays viewed in dark environments and to the mid- and high-frequency distortions represented in our dataset. They should therefore not be assumed to generalize to smaller displays, bright ambient environments, shorter viewing distances, or low-frequency and display-related distortions.
}

\edit{
The study also considered only two viewing distances, two display luminance levels, one OLED display, one codec and encoder preset, one upscaling method, and 16 short HDR contents without audio. The dim condition reduced the displayed content luminance by a factor of 8 while retaining a relatively high peak luminance (600 versus 700~cd/m$^2$), so it does not represent a display with lower peak luminance (e.g., an HDR 400 display).
Moreover, the methodology we employed to adapt the metrics to the viewing distance, using PPS rescaling should be interpreted as an engineering approximation within the investigated conditions rather than as a general model of visual sensitivity.
}

\edit{
Finally, the subjective task measured fidelity to a reference rather than overall preference or quality of experience, and did not account for streaming events such as stalling or bitrate switching. The reference videos were high-quality compressed versions because simultaneous playback of uncompressed 4K HDR streams was impractical; although they achieved more than 9.5~JOD according to ColorVideoVDP, a small residual compression effect cannot be excluded. Cross-luminance comparisons were also not collected, and the resulting scores were aligned using the common reference anchor described in \secref{subjective_scaling}. 
These limitations define the scope of the present dataset and motivate future work. Extending HDR-VDC to more viewing distances and luminance levels, ambient-light conditions, display sizes and peak-luminance capabilities, as well as additional codecs and distortion types, would allow the generality of the findings to be assessed more broadly. In particular, including low-frequency and display-related distortions would help test the limitations of PPS. Finally, end-to-end evaluation of viewing-condition-aware bitrate selection, including reliable client-side estimation of viewing conditions, would establish the practical streaming benefits suggested by our findings. Such extensions could also enable quality metrics to learn viewing-condition effects directly rather than relying on external pre-processing.
}



\section{Conclusions}\label{sec:conclusion}

Video quality metrics play a crucial role in optimizing video delivery. However, many existing metrics fail to account for the user's display and viewing conditions, reducing the effectiveness of the optimization pipeline. Hence, in this paper, we introduced a new HDR-VDC dataset. The dataset captures the visibility of streaming distortions in 4K HDR video content at two viewing distances and two display luminance levels.
The results from our subjective study indicate that both viewing distance and display luminance significantly impact the visibility of distortions, \edit{with viewing distance having a medium effect on quality and display luminance having a small effect within the range investigated}. Distortions are less visible at larger viewing distances or on dimmer displays. Additionally, video streaming can be further optimized when accounting for these factors. 
Furthermore, we investigated whether existing quality metrics could be adapted to account for both effects using PU21 encoding and PPS rescaling and if they could perform as well as metrics designed to natively account for these factors.
We evaluated 35 full-reference quality metrics on the dataset. We found that (a) metrics designed to account for the display and viewing conditions perform well in predicting their effect on quality, even when not calibrated on video streaming distortions; (b) Some quality metrics can be effectively adapted to account for this effect by employing PPS rescaling + PU21 encoding; however, many metrics cannot be adapted using such approach; and (c) most metrics do not show a good fit with the subjective scores, which indicates that there is room for improvement. We hope that our dataset can help to design and train future metrics.

In conclusion, this study emphasizes the importance of monitoring the viewing distance and display luminance in video quality studies. It also highlights the need for quality metrics that can natively account for these conditions, as a simple pipeline to adapt them to these conditions can not be guaranteed to perform well for each metric, or on all conditions.

\begin{acks}
We thank Mario Kleiner for his help in running the experiment in real time using the Psychtoolbox-3 framework. Additionally, we thank Maliha Ashraf and Joseph G. March for their insightful comments and advice throughout the project. Finally, we thank all participants of the experiment for their valuable time.
This work was supported by the Netflix Studentship to Dounia Hammou.
\end{acks}

\bibliographystyle{ACM-Reference-Format}
\bibliography{ref}

\newpage

\appendix

\section*{Online Appendix}

\begin{table*}[]
\centering
\caption{A comparison between our proposed HDR-VDC dataset and existing datasets in the literature. 
In the ``Test Methodology'' column, ``ACR'' refers to the absolute category rating methodology, ``PC'' to the pairwise comparison protocol,  ``DRM'' to distortion regions marking, ``DSIS'' to double stimulus impairment scale, and ``4AFC'' to the four-alternative forced-choice. 
}
\label{tab:datasets_comparison}
\resizebox{\textwidth}{!}{%
\begin{tabular}{l|cc|cc|cc|c|c}
\midrule
\textbf{Dataset} & \textbf{\begin{tabular}[c]{@{}c@{}}Video\\ Content\end{tabular}} & \textbf{\begin{tabular}[c]{@{}c@{}}HDR\\ Content\end{tabular}} & \textbf{\begin{tabular}[c]{@{}c@{}}Compression\\ Distortions\end{tabular}} & \textbf{\begin{tabular}[c]{@{}c@{}}Upscaling\\ Distortions\end{tabular}} & \textbf{\begin{tabular}[c]{@{}c@{}}Account for\\ Viewing Distance\end{tabular}} & \textbf{\begin{tabular}[c]{@{}c@{}}Account for\\ Display Luminance\end{tabular}} & \textbf{\begin{tabular}[c]{@{}c@{}}Publicly\\ Available\end{tabular}} & \textbf{\begin{tabular}[c]{@{}c@{}}Test\\ Methodology\end{tabular}} \\ \midrule
\citet{liu2014cid}        & \cellcolor[HTML]{FDDDD8}No                                       & \cellcolor[HTML]{FDDDD8}No                                     & \cellcolor[HTML]{CBE1CD}Yes                                                & \cellcolor[HTML]{FDDDD8}No                                               & \cellcolor[HTML]{CBE1CD}Yes                                                     & \cellcolor[HTML]{FDDDD8}No                                                       & \cellcolor[HTML]{CBE1CD}Yes                                           & ACR                                                                 \\ \midrule
\citet{amirpour2021impact}  & \cellcolor[HTML]{CBE1CD}Yes                                      & \cellcolor[HTML]{FDDDD8}No                                     & \cellcolor[HTML]{CBE1CD}Yes                                                & \cellcolor[HTML]{FDDDD8}No                                               & \cellcolor[HTML]{CBE1CD}Yes                                                     & \cellcolor[HTML]{FDDDD8}No                                                       & \cellcolor[HTML]{CBE1CD}Yes                                           & ACR                                                                 \\ 
\citet{sugito2022modeling}    & \cellcolor[HTML]{CBE1CD}Yes                                      & \cellcolor[HTML]{FDDDD8}No                                     & \cellcolor[HTML]{CBE1CD}Yes                                                & \cellcolor[HTML]{CBE1CD}Yes                                              & \cellcolor[HTML]{CBE1CD}Yes                                                     & \cellcolor[HTML]{FDDDD8}No                                                       & \cellcolor[HTML]{FDDDD8}No                                            & ACR                                                                 \\ 
\citet{kufa2019visual}      & \cellcolor[HTML]{CBE1CD}Yes                                      & \cellcolor[HTML]{FDDDD8}No                                     & \cellcolor[HTML]{CBE1CD}Yes                                                & \cellcolor[HTML]{CBE1CD}Yes                                              & \cellcolor[HTML]{CBE1CD}Yes                                                     & \cellcolor[HTML]{FDDDD8}No                                                       & \cellcolor[HTML]{CBE1CD}Yes                                           & PC                                                                  \\ \midrule
\citet{vcadik2011evaluation}      & \cellcolor[HTML]{CBE1CD}Yes                                      & \cellcolor[HTML]{CBE1CD}Yes                                    & \cellcolor[HTML]{CBE1CD}Yes                                                & \cellcolor[HTML]{FDDDD8}No                                               & \cellcolor[HTML]{FDDDD8}No                                                      & \cellcolor[HTML]{FDDDD8}No                                                       & \cellcolor[HTML]{CBE1CD}Yes                                           & DRM \\ 
\citet{azimi2018evaluating}     & \cellcolor[HTML]{CBE1CD}Yes                                      & \cellcolor[HTML]{CBE1CD}Yes                                    & \cellcolor[HTML]{CBE1CD}Yes                                                & \cellcolor[HTML]{FDDDD8}No                                               & \cellcolor[HTML]{FDDDD8}No                                                      & \cellcolor[HTML]{FDDDD8}No                                                       & \cellcolor[HTML]{FDDDD8}No                                            & DSIS                                                                 \\ \midrule
\citet{athar2019perceptual}     & \cellcolor[HTML]{CBE1CD}Yes                                      & \cellcolor[HTML]{CBE1CD}Yes                                    & \cellcolor[HTML]{CBE1CD}Yes                                                & \cellcolor[HTML]{FDDDD8}No                                              & \cellcolor[HTML]{FDDDD8}No                                                      & \cellcolor[HTML]{FDDDD8}No                                                       & \cellcolor[HTML]{FDDDD8}No                                            & ACR                                                                 \\ 
\citet{shang2022subjective}     & \cellcolor[HTML]{CBE1CD}Yes                                      & \cellcolor[HTML]{CBE1CD}Yes                                    & \cellcolor[HTML]{CBE1CD}Yes                                                & \cellcolor[HTML]{CBE1CD}Yes                                              & \cellcolor[HTML]{FDDDD8}No                                                      & \cellcolor[HTML]{FDDDD8}No                                                       & \cellcolor[HTML]{CBE1CD}Yes                                           & ACR                                                                 \\ \midrule
\citet{keller2024assessing}    & \cellcolor[HTML]{CBE1CD}Yes                                      & \cellcolor[HTML]{CBE1CD}Yes                                    & \cellcolor[HTML]{FDDDD8}No                                                 & \cellcolor[HTML]{CBE1CD}Yes                                              & \cellcolor[HTML]{CBE1CD}Yes                                                     & \cellcolor[HTML]{FDDDD8}No                                                       & \cellcolor[HTML]{FDDDD8}No                                            & PC                                                                  \\ \midrule
\citet{ebenezer2024hdr}    & \cellcolor[HTML]{CBE1CD}Yes                                      & \cellcolor[HTML]{CBE1CD}Yes                                    & \cellcolor[HTML]{CBE1CD}Yes                                                 & \cellcolor[HTML]{CBE1CD}Yes                                              & \cellcolor[HTML]{FDDDD8}No                                                     & \cellcolor[HTML]{CBE1CD}Yes                                                       & \cellcolor[HTML]{CBE1CD}Yes                                            & ACR                                                                  \\ \midrule
\citet{mikhailiuk2021effect} & \cellcolor[HTML]{FDDDD8}No                                       & \cellcolor[HTML]{FDDDD8}No                                     & \cellcolor[HTML]{CBE1CD}Yes                                                & \cellcolor[HTML]{FDDDD8}No                                               & \cellcolor[HTML]{CBE1CD}Yes                                                     & \cellcolor[HTML]{CBE1CD}Yes                                                      & \cellcolor[HTML]{CBE1CD}Yes                                           & \begin{tabular}[c]{@{}c@{}}QUEST\\ 4AFC\end{tabular}                \\ \midrule
HDR-VDC (ours)         & \cellcolor[HTML]{CBE1CD}Yes                                      & \cellcolor[HTML]{CBE1CD}Yes                                    & \cellcolor[HTML]{CBE1CD}Yes                                                & \cellcolor[HTML]{CBE1CD}Yes                                              & \cellcolor[HTML]{CBE1CD}Yes                                                     & \cellcolor[HTML]{CBE1CD}Yes                                                      & \cellcolor[HTML]{CBE1CD}Yes                                           & PC                                                                  \\ \midrule
\end{tabular}%
}
\end{table*}

\section{Video encoding parameters}\label{app:video_encoding_parameters}

The selected CRF values for each source sequence are reported in \tableref{selected_crf_values_v1}. Additionally, below we provide the command line used to introduce the distortions (Low-quality compression and a 1080p resolution) to the Bonfire sequence, as an example:

\begin{lstlisting}[language=bash]
$ ffmpeg -f rawvideo -pix_fmt yuv420p10le -s:v 3840x2160 -r 60 -colorspace bt2020nc -color_trc smpte2084 -color_primaries bt2020 -i Bonfire.yuv -pix_fmt yuv420p10le -vf scale=1920x1080:flags=lanczos -c:v libsvtav1 -crf 58 -preset 4 Bonfire_L_1920x1080.mp4
\end{lstlisting}

\begin{table*}[!hb]
\centering
\caption{CRF settings of the distorted videos. The same CRF value was used across all resolutions. The reference videos are the high-quality encoded videos (at their native resolution).
}
\label{tab:selected_crf_values_v1}
\resizebox{0.95\textwidth}{!}{%
\begin{tabular}{|l|cccccccccccccccc|}
\hline
\textbf{Source sequence index} & \textbf{1} & \textbf{2} & \textbf{3} & \textbf{4} & \textbf{5} & \textbf{6} & \textbf{7} & \textbf{8} & \textbf{9} & \textbf{10} & \textbf{11} & \textbf{12} & \textbf{13} & \textbf{14} & \textbf{15} & \textbf{16} \\ \hline
High-quality encoding CRF              & 7          & 9          & 9          & 5          & 9          & 9          & 9          & 9          & 7          & 7           & 11          & 7           & 9           & 7           & 7           & 7           \\ \hline
Medium-quality encoding CRF            & 43         & 37         & 43         & 33         & 35         & 41         & 49         & 37         & 43         & 33          & 45          & 27          & 37          & 37          & 36          & 55          \\ \hline
Low-quality encoding CRF               & 61         & 58         & 62         & 62         & 53         & 59         & 62         & 60         & 60         & 51          & 59          & 55          & 58          & 56          & 52          & 62          \\ \hline
\end{tabular}%
}
\end{table*}

\section{Content dependence of the viewing and display condition effects on quality}\label{app:effect_size_anova_vs_content}

\begin{figure}[]
    \centering
    \includegraphics[width=\linewidth]{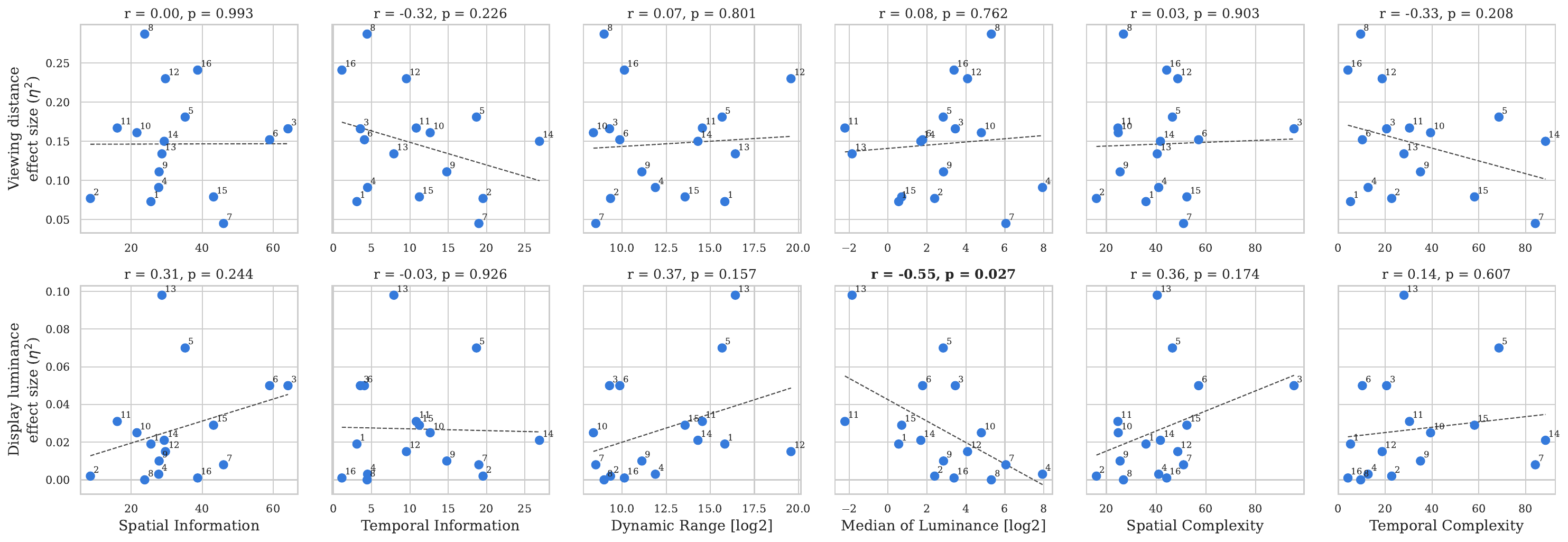}
    \caption{\edit{Per-content effect size ($\eta^2$) of the viewing distance (top) and the display luminance level (bottom) as a function of the content characteristics of the 16 reference videos, with the Pearson correlation coefficient ($r$), its $p$-value, and a least-squares fit (dashed line); Source indices correspond to the video indices in \figref{reference_screenshots}.}}
    \label{fig:effect_size_anova_vs_content}
\end{figure}

\edit{The ANOVA analysis reported in \secref{effect_viewing_display} shows that both viewing distance and display luminance level significantly affect perceived quality, with a substantially larger effect size for viewing distance. To examine whether these effects depend on the content, we correlated their effect sizes (reported in Section 2.2 of the supplementary) with the content characteristics described in \secref{source_sequences}: spatial and temporal information, dynamic range, median luminance, and spatial and temporal complexity. The results are shown in \figref{effect_size_anova_vs_content}. The effect size of viewing distance did not significantly correlate with any of the measured content characteristics ($|r| \leq 0.33$, $p > 0.2$), suggesting that its influence is largely independent of these content properties. Nonetheless, we observe a weak trend toward a smaller viewing-distance effect for contents with higher temporal information and complexity. One possible explanation is that highly dynamic contents already contain substantial temporal variation, which may reduce the relative contribution of viewing distance to the overall quality judgment. Interestingly, we do not observe a similar trend for spatial information or spatial complexity, suggesting that this relationship may be more specific to temporal characteristics than to content complexity in general. 
In contrast, the effect size of display luminance level was moderately and significantly correlated with median luminance ($r = -0.55$, $p = 0.027$): dimming had a larger impact on the quality of darker contents, such as SmithHammering and ShadowOfTheTombRaider, and a negligible impact on brighter contents, such as PorscheCloseUp and TrafficLight. This result is consistent with the expected effect of dimming on dark scenes, where distortions are shifted toward luminance levels at which contrast sensitivity is reduced. We also observe a weak trend toward a larger effect of display luminance level for contents with higher dynamic range. This may be because contents with a larger dynamic range contain a wider distribution of luminance levels, potentially making dimming more consequential for a larger portion of the content and shifting more details toward luminance levels with reduced contrast sensitivity. Given the small number of contents ($n=16$), these correlations should be considered exploratory, and the observed trends should not be interpreted as evidence of a causal relationship.
}

\section{Comparison with existing datasets} \label{app:comparison_existing_datasets}

It is necessary to analyze how the effect of the viewing distance and display luminance measured in our dataset differs from that measured in other available datasets. For this purpose, we analyze the effect size of the viewing distance and display luminance (when applicable) on three existing datasets, CID:IQ \cite{liu2014cid}, VCIP21 \cite{amirpour2021impact}, and VLIC \cite{mikhailiuk2021effect}.
The methodology to analyze the statistical significance of these factors was similar to the one we employed on our dataset to ensure a fair comparison. 
For the CID:IQ and VCIP21 datasets, we generated the distribution of the quality scores by resampling from a normal distribution $\mathcal{N}(\mu_i,\sigma_{i})$ , where $\mu_i$ is the reported quality score for condition $i$, and $\sigma_{i}$ is the standard error of the mean (as provided by the authors). We generated the distribution of the quality scores 1000 times (similar to the number of times we applied the 2-way ANOVA in our dataset), and we followed that by running the 1-way ANOVA on each generated distribution to calculate the effect size of the viewing distance on the quality scores. 
For the VLIC dataset, the authors measured the visually lossless threshold for each content. The visually lossless thresholds were measured for different viewing distances and display luminance levels, and all observers' results were reported in the dataset. Hence, we applied parametric resampling to the observers' scores using 5000 samples to generate the distribution of the mean thresholds. Using these distributions, we similarly applied the 2-way ANOVA to calculate the effect size of the viewing distance and display luminance on the visually lossless thresholds. 

We report the results in \tableref{comparison_datasets_effect_size}. We can observe from the table that while the VLIC dataset measured the effect of viewing distance on the visibility of distortions in SDR images, it shows a similar effect size to our HDR-VDC dataset. This indicates that the effect of the viewing distance is not dependent on the content type (SDR/HDR or Image/Video) since it will similarly impact the visibility of the distortions. In contrast, the CID:IQ and VCIP21 datasets showed smaller effect sizes. This is likely due to the higher sensitivity of the methodologies used in our and the VLIC datasets compared to the ACR methodology used in the their datasets. 
Interestingly, the VLIC dataset showed a much larger effect size for display luminance on distortion visibility compared to our dataset. This is because the VLIC dataset tested under significantly dimmer conditions, resulting in a larger effect size. These findings suggest that luminance levels can have an even greater impact on distortion visibility than reported in our dataset when tested under dimmer conditions.

\begin{table}[]
\centering
\caption{A summary of the datasets used in the comparison of the effect size of the viewing distance and display luminance on the visibility of distortions. 
}
\label{tab:comparison_datasets_effect_size}
\resizebox{0.8\columnwidth}{!}{%
\begin{tabular}{|l|c|c|c|c|c|c|}
\hline
\textbf{Dataset} & \textbf{\begin{tabular}[c]{@{}c@{}}Content\\ Type\end{tabular}} & \textbf{\begin{tabular}[c]{@{}c@{}}Test\\ Methodology\end{tabular}} & \textbf{\begin{tabular}[c]{@{}c@{}}Viewing\\ Distance\\ {[}\,ppd\,{]} \end{tabular}} & \textbf{\begin{tabular}[c]{@{}c@{}}Viewing\\ Distance\\ Effect Size\end{tabular}} & \textbf{\begin{tabular}[c]{@{}c@{}}Display\\ Luminance\\ {[}\cdms{]} \end{tabular}} & \textbf{\begin{tabular}[c]{@{}c@{}}Display\\ Luminance\\ Effect Size\end{tabular}} \\ \hline
CID:IQ \cite{liu2014cid}          & \begin{tabular}[c]{@{}c@{}}SDR\\ Images\end{tabular}            & ACR                                                                 & \begin{tabular}[c]{@{}c@{}}31.5\\ 63\end{tabular}                           & 0.035                                                                             & -                                                                                 & -                                                                                  \\ \hline
VCIP21 \cite{amirpour2021impact}          & \begin{tabular}[c]{@{}c@{}}SDR\\ Videos\end{tabular}            & ACR                                                                 & \begin{tabular}[c]{@{}c@{}}60\\ 92\\ 128\end{tabular}                       & 0.063                                                                             & -                                                                                 & -                                                                                  \\ \hline
VLIC \cite{mikhailiuk2021effect}            & \begin{tabular}[c]{@{}c@{}}SDR\\ Images\end{tabular}            & \begin{tabular}[c]{@{}c@{}}4AFC \&\\ QUEST\end{tabular}             & \begin{tabular}[c]{@{}c@{}}30\\ 60\end{tabular}                             & 0.109                                                                             & \begin{tabular}[c]{@{}c@{}}220\\ 10\end{tabular}                                  & 0.305                                                                              \\ \hline
HDR-VDC          & \begin{tabular}[c]{@{}c@{}}HDR\\ Videos\end{tabular}            & \begin{tabular}[c]{@{}c@{}}Pairwise\\ Comparison\end{tabular}       & \begin{tabular}[c]{@{}c@{}}60\\ 120\end{tabular}                            & 0.101                                                                             & \begin{tabular}[c]{@{}c@{}}700\\ 600\end{tabular}                                 & 0.015                                                                              \\ \hline
\end{tabular}%
}
\end{table}

\begin{figure}[!h]
    \centering
    \includegraphics[width=0.97\linewidth]{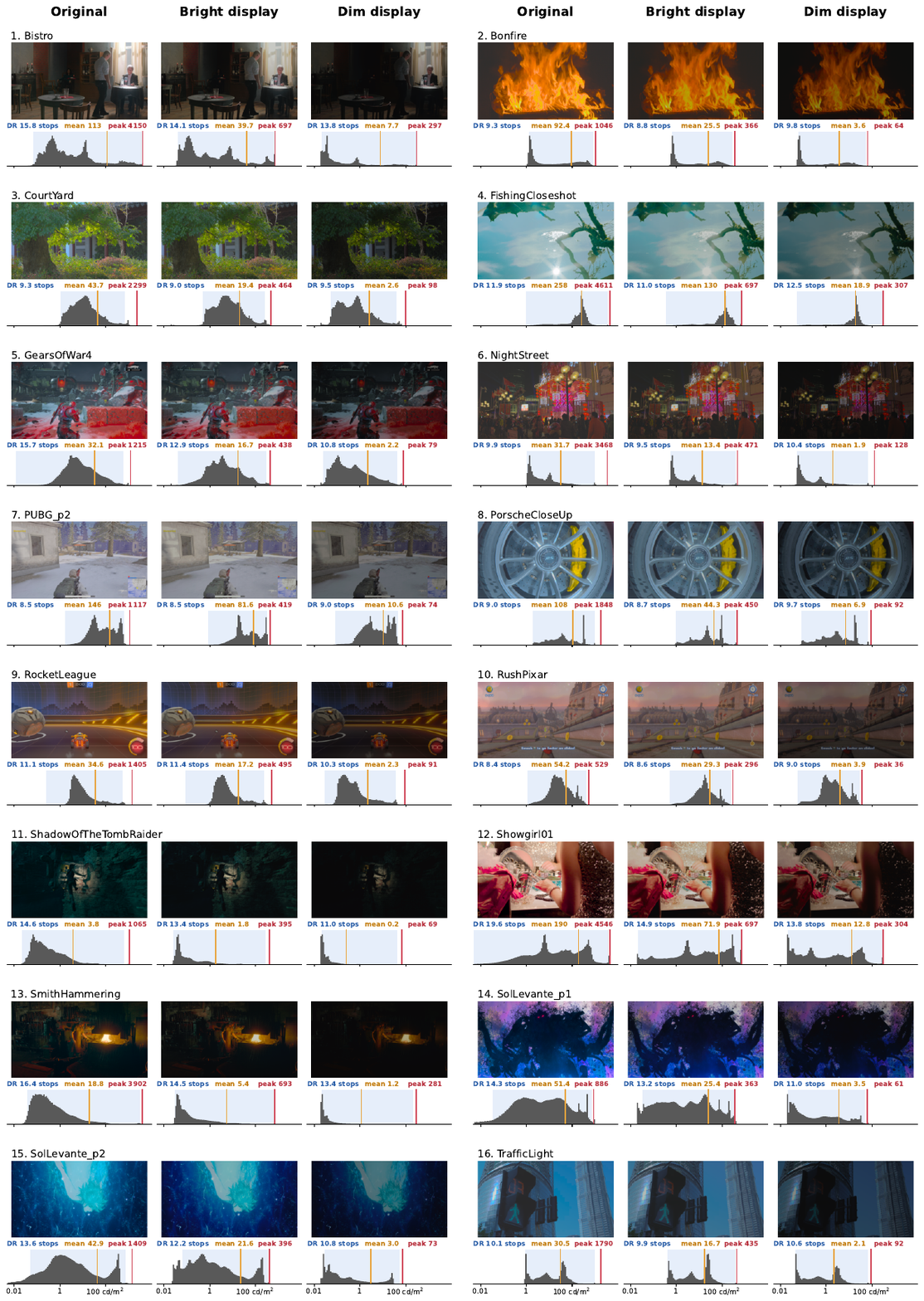}
    \caption{\edit{Representative frames of the 16 reference videos as encoded (Original) and as simulated on the LG OLED G2 in the bright and dim conditions, each with the luminance histogram of the whole video, its dynamic range (DR, blue band), mean luminance (orange) and peak luminance (red).}}
    \label{fig:reference_videos_display_model}
\end{figure}







\section{Scatter plots}\label{app:scatter_plots_figs}
\begin{figure}
    \centering
    \includegraphics[width=0.95\linewidth]{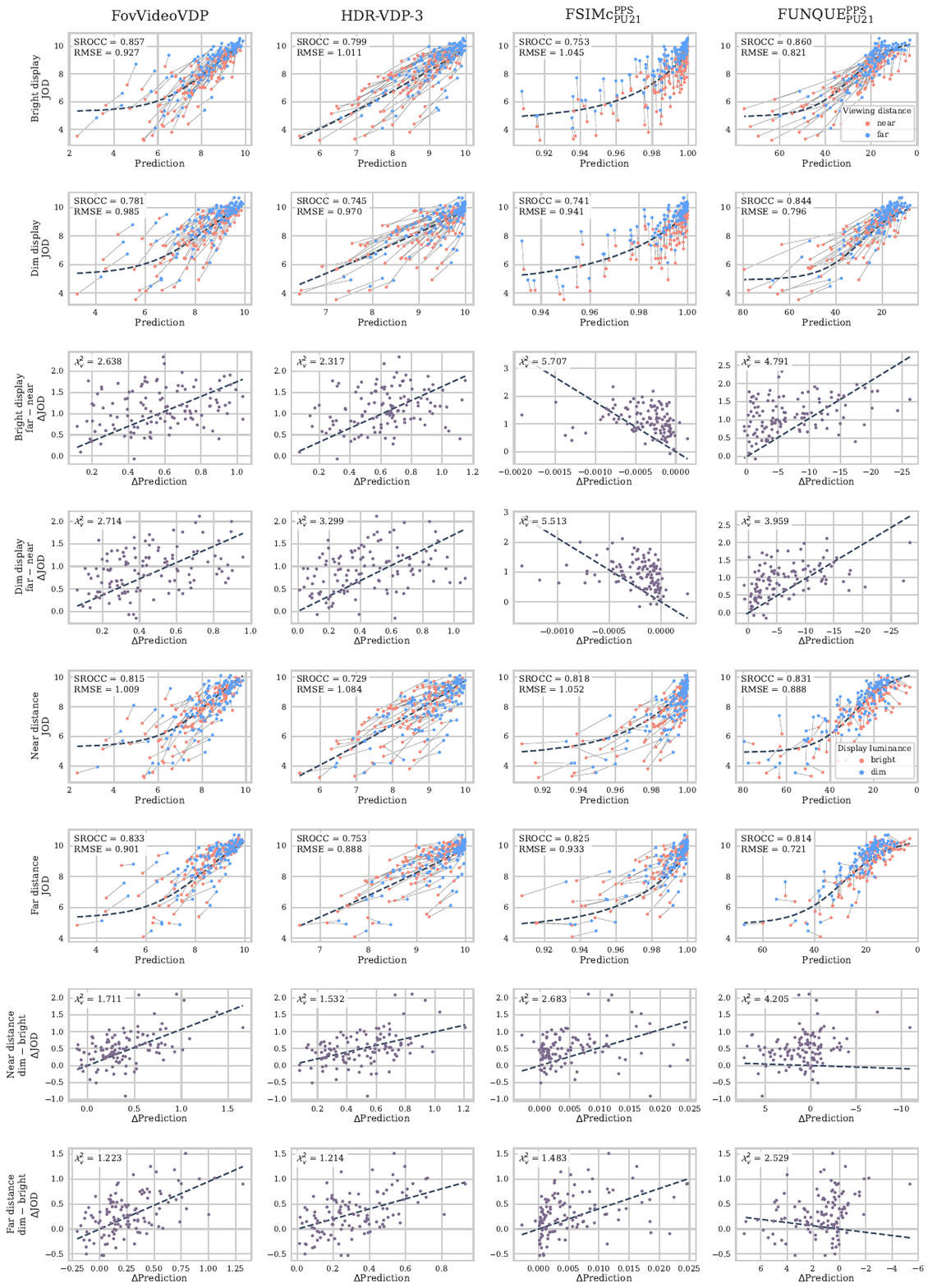}
    \caption{JOD scores versus FovVideoVDP, HDR-VDP-3, FSIMc, and FUNQUE predictions (columns). Rows 1–2: near and far viewing distance (same test videos connected) for the bright and dim display; rows 3–4: disentangled scores and predictions (far $-$ near). Rows 5–8: the same for display luminance (dim $-$ bright) at near and far viewing distance.}
    \label{fig:scatter_plots_all}
\end{figure}

In this section, we provide the scatter plots of some of the metrics across different viewing distances and display luminance levels in \figref{scatter_plots_all}. 
The figure shows, for each metric (column), the scatter plots of the JOD scores versus the metric predictions for each display luminance level (rows 1–2) and each viewing distance (rows 5–6). To investigate how well each metric accounts for the viewing conditions, we connect the same test videos viewed at different viewing distances (rows 1–2) or at different display luminance levels (rows 5–6). We also show the disentangled JOD scores versus the disentangled metric predictions for the viewing distance (far $-$ near, rows 3–4) and for the display luminance level (dim $-$ bright, rows 7–8). For metrics where a lower prediction indicates higher quality (FUNQUE), the horizontal axis is inverted. A metric that predicts the visibility of distortions well should produce predictions that correlate well with the JOD scores (rows 1–2 and 5–6). A metric that also predicts the effect of the viewing conditions should reproduce the change in JOD scores caused by each condition, so the lines connecting the same test videos should be parallel to each other. Equivalently, its disentangled predictions should align with the disentangled JOD scores (rows 3–4 and 7–8). 
From the figure, we can observe that metrics such as FovVideoVDP and HDR-VDP-3 can predict both the visibility of distortions and the effect of the viewing conditions. However, FSIMc cannot predict the effect of viewing distance at all, as it is rescaling the input frames, so it cannot be adapted with our pipeline. FUNQUE, on the other hand, its predictions correlate well with the JOD scores; however, it cannot predict well the effect of viewing conditions on the visibility of distortions after being adapted with our pipeline.

\end{document}